\documentstyle[12pt,a4]{article}

\begin{document}
\newcommand{\PRL}{{\em Phys. Rev. Lett. }}
\newcommand{\PRA}{{\em Phys. Rev. }}
\newcommand{\SOV}{{\em Sov. Phys. JETP }}
\newcommand{\NP}{{\em Nucl. Phys. }}
\newcommand{\JPB}{{\em Jour. of Phys. }}
\newcommand{\IJMP}{{\em Int. Jour. of Mod. Phys.  }}
\newcommand{\beq}{\begin{equation}}
\newcommand{\eeq}{\end{equation}}
\newcommand{\bea}{\begin{eqnarray}}
\newcommand{\eea}{\end{eqnarray}}
\newcommand{\hs}{\hspace{0.1cm}}
\newcommand{\spz}{\hspace{0.7cm}}
 \newcommand{\reset}{  \setcounter{equation}{0}  }
\newcommand{\th}{\theta}
\newcommand{\thb}{\bar{\theta}}
\newcommand{\ib}{ \bar{ \imath}}
\newcommand{\jb}{ \bar{ \jmath}}
\newcommand{\etab}{ \bar{ \eta}}
\newcommand{\kb}{ \bar{ k}}
\newcommand{\s}{\sigma}
\newcommand{\be}{\beta}
\newcommand{\Mrr}{I \!\! R} 
\newcommand{\xh}{ \hat{ x}}
\newcommand{\zb}{\bar{z}}
\newcommand{\zt}{\tilde{\zeta}}
\newcommand{\et}{\tilde{\zeta}}
\newcommand{\k}{\kappa}
\newcommand{\PL}{{\em Phys. Lett. }}
 \newcommand{\JMP}{{\em Jour. of Math. Phys.  }}
\newcommand{\CMP}{{\em Comm. Math. Phys.  }}
\setcounter{page}{0}
\topmargin 0pt
\oddsidemargin 5mm
\renewcommand{\thefootnote}{\fnsymbol{footnote}}
\newpage
\setcounter{page}{0}
\begin{titlepage}
\begin{flushright}
Berlin Sfb288 Preprint No. 201 \\
quant-ph/9604009\\
revised version
\end{flushright}
\vspace{0.5cm}
\begin{center}
{\Large {\bf On the absence of bound-state stabilization through short
ultra-intense fields} }

\vspace{1.8cm}
{\large A. Fring$^*$, V. Kostrykin$^{\dag}$ and R. Schrader$^*$}
\footnote{e-mail address: Fring or Schrader@omega.physik.fu-berlin.de,
Kostrykin@ilt.fhg.de}  \\
\vspace{0.5cm}
{\em $^*$ Institut f\"ur Theoretische Physik\\
Freie Universit\"at Berlin, Arnimallee 14, D-14195 Berlin, Germany\\
$^{\dag}$ Institut f\"ur Reine und Angewandte Mathematik\\
RWTH Aachen, Templergraben 55, D-52056 Aachen, Germany}

\end{center}
\vspace{1.2cm}
 
\renewcommand{\thefootnote}{\arabic{footnote}}
\setcounter{footnote}{0}
 
\begin{abstract}
We address the question of whether atomic bound states begin to stabilize
in the short ultra-intense field limit. We provide a general theory of
ionization probability and investigate its gauge invariance.
For a wide range of potentials we find an upper and
lower bound  by non-perturbative methods, which clearly exclude the 
possibility that the ultra intense
field might have a stabilizing effect on the atom. 
For short pulses we find almost complete
ionization as the field strength increases.
\end{abstract}
\vspace{.3cm}
\centerline{June 1996}
\end{titlepage}
\newpage

\renewcommand{\theequation}{\mbox{\arabic{section}.\arabic{equation}}}
\setcounter{equation}{0}

\section{Introduction}
Fermi's Golden Rule, as one of the central elements in quantum mechanics,
has served for many years for the understanding of photoionization rates
of atoms in weak radiation fields. Its origin is, however, perturbative
and therefore when applying very intense fields (with intensities
which are greater or of the order of one
atomic unit $3.5 \times 10^{16} {\rm W}/{\rm cm}^2$)
one leaves its range of validity. 
With the advance of laser
technology this high intensity region has become accessible
to real experiments in the form of laser pulses of 1ps or less,
at frequencies ranging from the infrared to the ultraviolet
\cite{Yaji}.
The prediction of atomic ionization rates are of practical importance
for instance in the study of gas breakdown \cite{Dew}. 

In order to treat the new regime, several alternative approximation
methods have been proposed. On one side \cite{Keld,Per,FaisalReiss} they are
based on a perturbation around the Gordon-Volkov
solution \cite{GordonVolk} of the Schr\"odinger equation. The question 
of convergence of these series
and their precise range of validity has not yet been put on firm grounds.
Despite these problems, these methods have been applied  to find
numerical solutions for the ionization probabilities. On the other side
there exist a vast number of numerical studies, which make use of 
numerical solutions of the Schr\"odinger equation, high-frequency 
approximations \cite{HFtheory} or  the Floquet 
approximation \cite{Floquet}.  Most computations have been carried out in 
one dimension  \cite{onedim,Gelt2}, in the hope that the essentials of 
the full three-dimensional physics are already present in this simplified
situation. There exist arguments which put them in question  
\cite{Reiss,Kul}, since in
comparison with the full three dimensional situation, they do not
account for the full angular  dependence and may provide
misleading results. Recently there have been full three dimensional 
computations \cite{Gelt1,Pont,Horbatsch,Kul,Lat,Casati}. But the complete
problem has not been solved yet and as it is pointed out in \cite{Gav},
``even the simplest one-electron atom in an intense laser field
presents too great a challenge for truly \it ab inito \rm
numerical work, and a variety of compromises have been developed". These
compromises are partly located in the numerical methods themselves, but
put partly  constraints onto the physics, such as for instance the 
introduction of mask functions or the approximation of the continuum by
cutting  of the high energetic states.

Several authors claim to have found the very surprising and 
counter-intuitive result, that the bound state of the atom stabilizes
as the field strength increases \cite{Pont,Horbatsch,Kul,Casati}.
Similar  results have also been obtained by many authors 
for the one dimensional  situation \cite{onedim}. In fact these findings
are so surprising that
``a dramatic shift in viewpoint is required to explain the physics
of atoms in very strong laser fields '' \cite{KulS}. 
We shall comment  on these results below and
for the moment refer the reader to the  review article on these findings
by Eberly and Kulander \cite{KulS} and one by Geltman
\cite{Gelt3}, who takes the opposite point of view that atomic bound states
do not stabilize as the field strength is increased and who asserts 
that the ``conventional interpretation of the theory of the interaction
of radiation and atoms is quite sound even in this regime''. The latter
point of view is also supported in \cite{nonstabil}.

Evidence for atomic-stabilization in superintense laser fields has 
also been obtained from the study of several classical dynamical
systems \cite{classical}.  

Up to now there exist no data for intensities of one atomic unit, i.e.
in the high intensity region for which the theoretical predictions are made,
such that the controversy could  be settled from the experimental side.
So far there  exist some experiments for lower intensities
$10^{13} W/cm^2$, which provide evidence for some sort of stabilization
\cite{experiments}.

The  controversy  is mainly based on numerical results and
a detailed theoretical analysis of the problem which involves analytic 
expressions only does not exist so far. The main intention 
of our paper is  to provide an alternative approach to the matter. 
We consider the Schr\"{o}dinger
equation for an atom in a linear polarized electric field,
\begin{displaymath}
i\frac{\partial\psi}{\partial t}=(-\Delta/2+V+z\cdot E(t))\psi,
\end{displaymath}
where $E(t)$ stands for the intensity of the field and is
supposed to have finite duration 
(for instance, 1 ps = 4.17$\times 10^4$ a.u.).
We do not specify $E(t)$ in more detail: it can be, for instance,
a pulse which contains a number of optical periods (the
frequency of which is determined by the frequency of the laser) 
possibly with some turn-on and turn-off parts. Such kind of pulses were used
in the search for  possible suppression of ionization. We note that
1 ps pulses have a duration comparable with a classical Kepler
period for the highly excited Rydberg states ($T_K=2\pi n^3$).
Another example are half-cycle pulses with duration of about 500 fs,
generated in the experiments of Jones et al. \cite{Jones}.   
However, the maximal intensity reached in these experiments was about
$10^{-6}$ a.u., so that these pulses are ultimately far from
the ultra-intense limit.

We suppose furthermore that the wave function $\psi(\vec{x},t)$ 
is given by the bound state wave function of $-\Delta/2+V$ before the
pulse is turned on. One can easily estimate
that relativistic effects might be appreciable as soon as
$E(t)$ is so strong that the classical theory
predicts electron velocities approaching
the speed of  light, or more precisely when in atomic units
the electric field strength time the frequency is of the order of the
fine structure constant.
According to the estimates in \cite{Gav}
this occurs for typical frequencies for laser 
intensities $E\approx 10^{18}{\rm W}/ {\rm cm}^2$. 

Our results below show that atoms do not become resistant
to ionization when exposed to short ultra-intense laser pulses.
Our statements are of 
qualitative nature, in the sense that they provide upper and lower 
bounds and do not predict precise values of the ionization probabilities.
The methods we use cover all possible pulses, i.e. also those which
are very popular in the literature with smoooth turn on and off. Our
arguments cover all frequency regimes, including the high frequencies 
for which stabilization is supposed to occur. For pulses which are
not switched on smoothly, our results typically hold for very short times
of the order of one atomic unit. With a smooth switch on of the pulse
one may extend the region of validity. We provide expressions
for two upper bounds, (3.2) and (3.14) valid in the region when
$ \left(\int_0^{\tau} E(t) dt  \right)^2 /2$ (the classical energy transfer
of the pulse)  is smaller than the binding energy and the other 
valid without restriction. The lower bound holds when 
$ \left(\int_0^{\tau} E(t) dt  \right)^2 /2  > -E$.

The paper is organized as follows. In Section 2 we formulate a
general theory of ionization probability and prove its gauge 
invariance. We also make contact to the various approximation methods
based on perturbative expansions. In Section 3 we briefly discuss
these methods in the context of  quantum mechanical one-particle
Stark Hamiltonians
and provide the proofs for the upper and lower bounds
for the ionization probability for a wide range of one-particle 
potentials, which in particular include
all potentials appearing in 
atomic and molecular physics. 
In Section 4 we 
state our conclusions. In Appendix A we provide an upper bound for
the Coulomb potential and in Appendix B we optimize this bound for the
ground state of the hydrogen atom.   

\section{The general theory of ionization probability and its gauge invariance}

In this section we will give a general discussion of the ionization probability
and its gauge invariance. Gauge invariance is of course necessary for
observable quantities and it is conventional wisdom for the case of the 
ionization probability. However, we were not able to locate an explicit 
reference with a proof and will therefore include a discussion on this issue.
We will relate our arguments to familiar concepts in scattering 
theory and explicitly discuss its relevance in the context of the 
Stark Hamiltonian.
In the last part of this section we will show the gauge covariance of 
time-dependent perturbation theory. 
In order to convey the general ideas 
we will avoid bulky mathematical notations in this section. 

Let $H(t) (-\infty <t<\infty)$ be a general
time dependent selfadjoint Hamiltonian in some Hilbert space $\cal H$ and let
$U(t,t')$ denote the resulting time evolution operator from $t'$ to $t$, i.e.
$U(t,t')$ satisfies
\begin{eqnarray}\label{2.1}
i\partial_tU(t,t') &=& H(t) U(t,t')\nonumber\\
U(t,t')U(t',t'') &=& U(t,t'')\\
U(t,t) &=& {\bf 1} \nonumber
\end{eqnarray}
for all $t,t',t''$. In the context of the Stark Hamiltonians, $U(t,t')$
exists for all $t,t'$ and is unitary (see below).

Assume now that $H(t)$ approaches an operator $H_{+ }$ 
for $t\to\infty$ and $H_{- }$ for $t\to -\infty$, i.e.
\begin{eqnarray}\label{2.2}
H_{+ } &=& \lim_{t\to\infty} H(t)\nonumber\\
H_{- } &=& \lim_{t\to -\infty} H(t)
\end{eqnarray}
holds in a suitable sense. It is important to note that we do not assume
$H_{+ }$ to equal $H_{- }$. In fact, for the
Stark Hamiltonian in certain gauges, these operators will in general differ
(see below).
In analogy to the scattering matrix (see below) we define the abstract S-Matrix
to be the following weak limit, i.e. the limit for matrix elements
(if it exists)
\begin{equation}\label{2.3}
S=\lim_{t\to +\infty \atop t'\to -\infty} \exp~it H_+  \cdot U(t,t') \cdot 
\exp -it' H_{- }.
\end{equation}
In particular $S$ exists trivially and is unitary if $H(t)$ becomes
stationary for all large $|t|$, i.e. if $H(t) =H_{+ }$ for all $t\ge t_+$ and $H(t)=H_{- }$
for all $t\le t_-$ for suitable finite $t_-,t_+$.  
In this case we will call $H(t)$ a {\it finitely pulsed
Hamiltonian}. $S$ then takes the form
\begin{equation}\label{2.4}
S=\exp ~it H_{+ }\cdot U(t,t')\cdot \exp -it' H_{- }
\end{equation}
for all $t\ge t_+$ and all $t'\le t_-$. In particular S is then unitary.

Let $P_{+ }$ be the orthogonal projection onto the subspace spanned by
the bound states of $H_{+ }$. $P_{- }$ is defined analogously in
terms of $H_{- }$. Then for any normalized state $\psi$ in the range of
$P_{- }$ its ionization probability is defined to be
\begin{equation}\label{2.5}
I(\psi) =\|({\bf 1}-P_{+ })S\psi\|^2.
\end{equation}
Here $ \| \psi \| $ denotes the Hilbert space norm, i.e. 
$ \| \psi \|^2 = \langle \psi, \psi  \rangle $
For the case when $H_{+ } =H_{- }$ this agrees with the definition
used in \cite{EKS,KS1,KS2}. For a finitely pulsed Hamiltonian
we have
\begin{equation}\label{2.6}
I(\psi) =\|({\bf 1}-P_{+ }) U(t_+,t_-)\psi \|^2
\end{equation}
whenever $\psi$ is a bound state of $H_{- }$. 

Abstract gauge transformations
are now introduced as follows. Let $A(t)$ $(-\infty$ 
$ <t<\infty)$ be a one parameter family of
unitary operators (suitably differentiable in $t$). If $\psi(t)$ is a solution
of the Schr\"odinger equation
$$
i\partial_t \psi(t) = H(t)\psi(t)
$$
then $\psi' (t) = A(t) \psi(t)$ is a solution of the equation
\begin{eqnarray}
i\partial_t\psi'(t) &=& i\left( \partial_tA(t) \right) 
\psi(t)+A(t) i\partial_t \psi(t)
\nonumber\\
&= &i(\partial _tA(t))A(t)^{-1} \psi' (t)+A(t)H(t)A(t)^{-1} \psi'(t)
\label{2.7}\\
&=& H'(t)\psi'(t) \nonumber
\end{eqnarray}
with
\begin{equation}\label{2.8}
H'(t) = i(\partial_tA(t) )A(t)^{-1} +A(t) H(t) A(t)^{-1}
\end{equation}
being formally selfadjoint. If $U'(t,t')$ is the time evolution operator for
$H'(t)$ then obviously 
\begin{equation}\label{2.9}
U'(t,t') =A(t) U(t,t')A(t')^{-1}.
\end{equation}
We note that the set of all gauge transformations form a non-commutative group
under the obvious multiplication rule $(A_1A_2)(t) =A_1(t) A_2(t)$, with unit
${\bf 1} (t) ={\bf 1}$ and inverse $A^{-1} (t) =A(t)^{-1}$. The familiar
interaction picture used in scattering theory is now a special case.
Indeed, to be more specific assume $H(t)$ to be of the form
$H(t) =H_0 +H_I(t)$, where $H_0$ is the ``free" Hamiltonian and $H_I(t)$ the
(possibly time dependent) interaction Hamiltonian. Set 
\begin{equation}\label{2.10}
A(t) =\exp  it H_0  .
\end{equation}
Then
\begin{equation}\label{2.11}
H'(t) = H_I' (t)
\end{equation}
is the Hamiltonian in the interaction picture with
\begin{equation}\label{2.12}
H_I'(t) =\exp\; it H_0 \cdot H_I(t)\cdot 
\exp-it H_0 .
\end{equation}
In this case the limit (if it exists)
\begin{equation}\label{2.13}
S(H,H_0) = \lim_{t\to+\infty\atop t'\to-\infty}  U'(t,t')
= \lim_{t\to+\infty\atop t'\to-\infty} \exp~it H_0\cdot U(t,t')\cdot \exp-it'H_0
\end{equation}
is called the scattering matrix (S-matrix) for the pair $(H,H_0)$.
Let us elaborate briefly in what sense there is an analogy between the quantity
(2.3) and the case of ordinary potential scattering, i.e.
where $H_0$ is $-\Delta/2$ and $H_I(t) =V$ is a potential. In the latter case 
$H=H_0 +V$ is
compared with $H_0$ in spatial regions far out, i.e. where $V$ is small
and where the incoming and outgoing wave packets are located for large $|t|$. 
In the  case we are interested in presently,
$H(t)$ is compared with $H_{+ }$ for large positive times and with 
$H_{- }$ for large negative times. To summarize: In one case one 
compares Hamiltonians 
for large spatial coordinates and in the other case for large time coordinates.

To apply this general concept of gauge transformations and gauge covariance
to our discussion of ionization, assume now in addition that $A(t)$ approaches
suitably unitary operators $A_{+ }$ and $A_{- }$ when $t\to +\infty$
and $t\to-\infty$, respectively.

Then $H'(t)$ (see (2.8)) approaches
\begin{equation}\label{2.14}
H'_{+ } =A_{+ } H_{+ } A^{-1}_{+ }
\end{equation}
and
\begin{equation}\label{2.15}
H'_{- } =A_{- } H_{- } A_{- }^{-1}
\end{equation}
as $t\to +\infty$ and $t\to-\infty$ respectively.

By formal manipulations we therefore have 
\begin{equation}\label{2.16}
S' =A_{+ } S A_{- }^{-1}.
\end{equation}
In particular if $H(t)$ is finitely pulsed and if in addition $A(t)$ is
stationary for all large $|t|$, then $H'(t)$ is also finitely pulsed, $S'$
exists, is unitary  and (2.16) holds.

In general, by (2.14) and (2.15), if $P'_{\pm }$ are the orthogonal
projections onto the space spanned by the bound states of $H'_{\pm }$,
we have
\begin{equation}\label{2.17}
P'_{\pm } \;\; = \;\; A_{\pm } \; P_{\pm } \;A_{\pm }^{-1}.
\end{equation}
In particular $A_{- }\psi$ is in the range of $P'_{- }$ if $\psi$
is in the range of $P_{- }$. Inserting (2.16) and (2.17) gives the
desired gauge invariance in the form
\begin{eqnarray}\label{2.18}
 I' (A_{- } \psi) &=&\| ({\bf 1}-P_{+ }')S' A_{- } \psi\|^2\nonumber\\
&=&\|A_{+ } ({\bf 1}-P_{+ }) A_{+ }^{-1} A_{+ } SA_{- }^{-1}
A_{- } \psi\|^2 \\
&=&\|({\bf 1}-P_{+ })S\psi \|^2 =I(\psi), \nonumber
\end{eqnarray}
since by assumption $A_{+ }$ is unitary. In the example we will be
interested in $A_{- } ={\bf 1}$. In this case $H'_{- } =
H_{- }$ and (2.18) takes the simpler form $I'(\psi) =I(\psi)$.
From the proof we see that gauge invariance is an important regulating
principle in the following sense. One has to choose the projection $P_{+ }$
in (2.5) and not $P_{- }$ in order to obtain gauge invariance.

We now apply these concepts to the theory of the time dependent Stark
Hamiltonian. In order to make notations more transparent we choose
a linearly polarized electric field, which does however not limitate
our discussion since more general fields 
may be simply obtained by replacing $z \rightarrow
\vec{x}$ and $ E(t) \rightarrow \vec{E}(t)$ ($\vec{x} \in \Mrr^3,
\vec{E} \in \Mrr^3$), such that particular other choices, like 
for instance, circular polarized light may be easily be derived from there.
We do, however, assume a dipole
approximation, such that the electric field becomes a function only
of time and thus is independent of space. Then using atomic units
$ \hbar=e=m_e=c \cdot \alpha=1$ 
we consider on the Hilbert space $L^2(\Mrr^3,d^3 x)$ 
the 3 time-dependent Hamiltonians
\begin{eqnarray}\label{2.19}
H_1(t) &=& - {\Delta\over 2} +V +z \cdot E(t)\nonumber\\
H_2(t) &=& {1\over 2} (-i\nabla -b(t) e_z)^2 +V \\
H_3(t) &=& -{\Delta\over 2} +V(\vec{x} -c(t) e_z) \;\; \; . \nonumber
\end{eqnarray}
Here $V$ is an arbitrary potential. $e_z$ is the unit vector in the 
$z$-direction and $V(\vec{x} -\vec y)$ is the shifted potential, i.e.
the multiplication operator 
on wave functions given as
$(V(\vec{x} -\vec y)\psi) (\vec x)=V(\vec x-\vec y)\psi (\vec x)$.
Also $E(t)$ is the electric field, assumed to vanish unless $0\le t\le\tau$
(i.e. $t_-=0$ and $t_+ =\tau>0$ in the notation above). 
Apart from this condition the pulse $E(t)$ may be arbitrary. We only make the
mathematical restriction, that $E(t)$ is piecewise continuous, which means
that the pulse may have jumps and all commonly used  enveloping shapes, for
instance cosine squared, smooth adiabatic turn on and off, etc.,
are included. Then the following quantities $a(t),b(t)$ and
$c(t)$ are well defined  
\bea 
b(t) &=& \int^t_0 E(s) ds\\
c(t) &=& \int^t_0 b(s) ds\;\; =\;\;  t b(t) - \int_0^t s E(s) ds 
\label{cterm}\\
a(t) &=& \frac{1}{2}  \int^t_0 b(s)^2 ds\;\; .
\eea
Note that $b(\tau)e_z$ describes the classical momentum transfer of the pulse,
such that ${1\over 2}b(\tau)^2$ is the classical energy transfer. Also
$c(\tau)c_z$ is the classical displacement caused by the pulse.
Then $H_2(t)$ is obtained from $H_1(t)$ by the gauge transformation 
$$
A_{2\leftarrow1}(t) = \exp~ib(t) z. 
$$
$H_1(t)$ is obtained from $H_3(t)$ by the Kramers-Henneberger transformation
\cite{K,H,CFKS}. 
\begin{equation}\label{2.23}
A_{1\leftarrow3}(t)=T(t)=\exp -ia(t) \cdot \exp -ib(t) z \cdot  \exp ic(t)p_z.
\end{equation}
Therefore we call $H_3(t)$ the Hamiltonian in the Kramers-Henneberger gauge.
We note that a corresponding transformation in Quantum 
Electrodynamics was introduced by Pauli and Fierz already in 1938 but 
with a different motivation \cite {PF}. The Hamiltonian $H_1(t)$ is usually
referred to as the Hamiltonian in the length- or electric field gauge, 
whereas $H_2(t)$ is denoted as the Hamiltonian in the velocity-, radiation- 
or Coulomb gauge.

As a consequence $H_2(t)$ is obtained from $H_3 (t)$ by  the gauge transformation
\begin{equation}\label{2.24}
A_{2\leftarrow 3}(t) \;\;= A_{2\leftarrow1} (t) A_{1\leftarrow 3}(t)
=\exp-ia(t) \cdot \exp  ic(t)p_z.
\end{equation}
The general gauge transformation $A_{j\leftarrow i}(t)$ for $H_i(t)\to H_j(t)$
is then obtained from the rules
\begin{eqnarray}\label{2.25}
&&A_{j\leftarrow i} (t) = A_{i\leftarrow j}(t)^{-1}\nonumber\\
&& A_{j\leftarrow k}(t) = A_{j\leftarrow i} (t) A_{i\leftarrow k}(t).
\end{eqnarray}
$H_1(t)$ and $H_2(t)$ are finitely pulsed. Note however that in general
$H_3(t)$ is {\it not} finitely pulsed and hence has not always a
proper limit as 
$t\to +\infty$. This is due to the fact that $b(t)$ is constant for
$t\ge\tau$ such that $c(t)$ grows linearly in   $t\ge\tau$ whenever 
$b(\tau)\not= 0$, as is apparent from (\ref{cterm}). Nevertheless the Kramers-Henneberger gauge is quite useful
as we shall see below.
These observations are related to the fact that
$A_{2\leftarrow1}(t)$ becomes stationary for large $t$ but in general
not $A_{2\leftarrow3}(t)$ and $A_{1\leftarrow3}(t)$.
Note that by assumption on $E(t),A_{j\leftarrow i}(-\infty) ={\bf 1}$ and thus
\begin{equation}\label{2.26}
H_{1,+ } =H_{1,- } =H_{2,- }  
=H_{3,- }=H_0+V.
\end{equation}
However, in general $H_{2,+ } \not= H_{2,- }$.
 
The case $V\equiv 0$ is of special interest, since it corresponds
to the situation in which the Schr\"odinger equation admits an exact 
solution, which is usually referred to as the Gordon-Volkov solution
\cite{GordonVolk}. Call the resulting operators
$H_{0,i}(t)(i=1,2,3)$, such that in particular $H_{0,3}(t)=-\Delta/2$.
The kernels of the resulting time evolution operators $U_{0,i}(t,t')$ can be
calculated explicitly. Indeed, we start from the familiar relation
for the free particle evolution operator (see e.g.  \cite{GW})
\begin{eqnarray}\label{2.27}
U_{0,3}(\vec{x},t;\vec{x}',t')&=&\langle \vec{x}|U_{0,3}(t,t')|
\vec{x}'\rangle
\; = \; \langle \vec{x}|\exp i(t-t'){\Delta\over 2}|\vec{x}'\rangle \\
&=&{1\over (2\pi i(t-t'))^{3/2}}
\exp i{(\vec{x}-\vec{x}')^2\over 2(t-t')}.
\end{eqnarray}
Obviously
\bea
\langle \vec{x}|\exp~i b(t)z|\vec{x}'\rangle &=&
\exp~ib(t)z \cdot \delta^3(\vec{x}-\vec{x}') \nonumber \\
\langle \vec{x}|\exp~i a(t)|\vec{x}'\rangle &=&
\exp~i a(t) \cdot \delta^3(\vec{x}-\vec{x}') \nonumber
\eea
and
$$\langle \vec{x}|\exp ic(t) p_z|\vec{x}'\rangle 
=\langle \vec{x} - c(t)e_z,\vec{x}' \rangle =
\delta^3 (\vec{x}-\vec{x}'-c(t) e_z)$$
such that
\begin{eqnarray}\label{2.28}
\langle \vec{x}|T(t)|\vec{x}'\rangle 
&=& \overline{ \langle \vec{x}'|T(t)^{-1}|\vec{x}\rangle } \nonumber\\
&=& \exp-ia(t) \cdot \exp-ib(t)z \cdot \delta^3(\vec{x}-\vec{x}'-c(t)e_z).
\end{eqnarray}
By (2.9) we immediately obtain the following (again well known) relations
(see e.g. \cite{Dav,BeckerDav,GordonVolk,Gelt1}).
\begin{eqnarray}\label{2.29}
U_{0,1} (\vec{x},t;\vec{x}',t') 
&=& {1\over (2\pi i(t-t'))^{3/2}} \exp i(a(t')-a(t))\nonumber\\
&&\times \exp i(b(t')z-b(t)z') \exp i \frac{(\vec{x}-c(t)e_z-\vec{x}'
+c(t')e_z)^2}
{2(t-t')} \nonumber\\
U_{0,2} (\vec{x},t;\vec{x}',t') 
&=& {1\over (2\pi i(t-t'))^{3/2}} \exp i(a(t')-a(t))\nonumber\\
&&\times \exp i \frac{(\vec{x}-c(t)e_z-\vec{x}'
+c(t')e_z)^2} {2(t-t')}.  \nonumber
\end{eqnarray}
The kernel of $U_{0,1}(t,t')$ is often called the Gordon-Volkov propagator
(see e.g. \cite{GordonVolk,Dav,Gelt1}).

We finally give a discussion of time dependent perturbation theory and its gauge
covariance. Returning to the general set-up, let $H(t)$ be a `` perturbation" of
$K(t)$. If $W(t,t')$ denotes the time evolution for $K(t)$, we have the 
generalized Du Hamel's formula (see e.g. \cite{RS}) in the  form
\begin{eqnarray}\label{2.30}
U(t,t')&=& W(t,t') -\int^t_{t'} {d\over ds} \left[ U(t,s) W(s,t')\right]ds\nonumber\\
&=& W(t,t') -i \int^t_{t'} U(t,s)\left[ H(s)-K(s)\right] W(s,t')ds.
\end{eqnarray}
We recall that in terms of the sometimes more familiar Green's function
$$G_{H } (t,t') =-i U(t,t') \theta (t-t') \;\; , $$
which satisfies
$$(i\partial_t-H(t))G_{H }(t,t') =\delta (t-t') {\bf 1}\; ,$$
relation (2.30) takes the form
\begin{equation}\label{2.31}
G_{H } (t,t') =G_{K } (t,t') 
+\int\limits^{+\infty}_{-\infty} G_{H }
(t,s) \left[ H(s)-K(s)\right] G_{K } (s,t')ds.
\end{equation}
Similarly one derives the relation
$$U(t,t') =W(t,t')-i \int^t_{t'} W(t,s) \left[ H(s)-K(s)\right] U(s,t')ds
\eqno(2.30')$$
and hence
$$G_{H } (t,t') =G_{K }(t,t') 
+\int\limits_{-\infty}^{+\infty} G_{K }
(t,s)\left[ H(s)-K(s)\right] G_{H } (s,t') ds. \eqno(2.31')$$
We recall that Du Hamel's formula in the form (2.31) and ($2.31'$) is the time
dependent version of the Lippmann-Schwinger equation (see e.g.\cite{Feyn,GW}) 
in the
case where both $H(s)$ and $K(s)$ are actually time-independent. Indeed, the
Lippmann-Schwinger equation may be obtained from these relations by taking
Laplace transforms.

The equations (2.30), (2.30'), (2.31) and (2.31') may be iterated by introducing
the left hand side into the right hand side, resulting in a ``power series 
expansion" of $U(t,t')$  in powers of $H(s)-K(s)$
and involving $W$ of the form
\begin{equation}\label{2.32}
U(t,t') =\sum_{n=0}^\infty U_n(t,t')
\end{equation}
with $U_0(t,t') =W(t,t')$ and with an analogous expansion for 
 $G_{H }$.
 Let us consider what happens under gauge transformations.
Du Hamel's formula is compatible with gauge transformations in the sense that
the relation
\begin{equation}\label{2.33}
U'(t,t') =W'(t,t') -i \int_{t'}^t U'(t,s) \left[H'(s)-K'(s)\right] W'(s,t')ds
\end{equation}
either follows directly for the pair $H'(t),K'(t)$ or by applying the
gauge transformation to (2.30) and using (2.8) and (2.9). This implies in
particular that
\begin{equation}\label{2.34}
U'_n (t,t') =A(t) U_n(t,t') A(t')^{-1}
\end{equation}
for all $n$. This applies in particular to the choices $H(t) =H_i(t)$
and $K(t)=H_{0,i}(t)$  given by (2.19) and the gauge transformations which
relate them.
However, some of the approximation methods used  for high
intensities, on which we shall comment more below,   
use the fact that one can decompose the Stark Hamiltonian in two
different ways, that is either treating the potential or the term
related to the electric field as a perturbation. 
Hence one obtains two versions for (2.30), which one may
combine iteratively. The series generated in this manner in
general does not respect gauge invariance order by order. A discussion of
this problem and a remedy for restoring gauge invariance
by including some terms of  next order, thus leading to cancelations,
may be found in \cite{Dav,Milonni}.

In the case of the interaction picture (see above) and with the choice
$K(t)=H_0$ such that $K'(t)=0$, the iteration of (2.33) in powers of
$H_I'(t)=H'(t)-K'(t)$ is just the famous Dyson series of the S-Matrix
$S(H ,H_0)$ in the limit $t\to +\infty,t'\to-\infty$. In the time
independent context of the Lippmann-Schwinger equation (see above) this corresponds
to iterating the Lippmann-Schwinger equation to obtain the Born series for the
S-Matrix (see e.g. \cite{GW}).
In the context we are presently interested in, such series expansions for the 
time evolution operator of finitely
pulsed Hamiltonians lead to a series expansion for the ionization probability when inserted
in (2.6). In the next section we will discuss the various approaches used so 
far
for the Hamiltonian $H_1(t)$ and its gauge transforms $H_2(t)$ and $H_3(t)$
given in (2.19).

\setcounter{equation}{0}

\section{Ionization of atoms in strong, short electric fields}

Using the notions of the previous section we start with a review and comparison
of methods and results obtained by previous authors.
Then we relate this in a first step to a new, rigorous upper bound on
the ionization probability, valid for all small $\tau$ and small classical momentum
transfer $b(\tau)$ and small displacement $c(\tau)$ (see 
the upper bound  below).
This result is also compared with another rigorous upper bound previously
obtained by  two of the authors 
(V.K., R.S.) \cite{KS1} as well as with a
result obtained  using a time-energy uncertainty relation given by
Pfeifer \cite{P}.
Secondly we prove a lower bound below, valid for all small
$\tau$ and all large $b(\tau)$, which in particular proves the absence
of stabilization.

Taking $H(t)=H_0+V+z\cdot E(t)=H_1(t)$ and $K(t)=H_0+V$ in (2.30), the resulting
perturbation series in the time dependent interaction $H(t)-K(t)=zE(t)$ for
$U_1(t,t')$, the time evolution operator for $H_1(t)$,
has been used by Lambropolous \cite{Lam}. Certainly for high
intensities $E(t)$ this is very problematic, since one requires several
terms in the expansion to achieve a reasonable result.

A more promising approach has been advocated by Perelomov, Popov and
Terentev \cite{Per},
who took $H(t)=H_0+V+z\cdot E(t)=H_1(t)$ and $K(t)=H_0+z\cdot E(t)=H_{0,1}(t)$.
Then $H(t)-K(t)=V$, thus leading to a power series expansion in $V$. Whenever
$|E|>|V|$ this seems to be very suggestive approximation and is based on the
fact that the time evolution operator $U_{0,1}(t,t')$ for $H_{0,1}(t)$ is the
Gordon-Volkov solution, which is known exactly (see equation after (2.29)). 
By the discussion of section 2, in the Kramers-Henneberger gauge this 
corresponds to a power series expansion in suitable translates of $V$.

A combination of these two methods has been proposed earlier in a seminal
paper on the subject by Keldysh 
\cite{Keld}, who
took the series for $U_1(t,t')$ with $z\cdot E(t)$ as a perturbation, but in the 
second iteration step inserted the time evolution operator $U_{0,1}$ instead
of the time evolution operator for $H_0+V$. In fact, it was demonstrated by
Davidovich et al. \cite{Dav} that to first order the Keldysh approximation and the
one of Perelomov et al. precisely coincide. 
When carrying out the same steps in the velocity gauge, i.e. for
$H(t) = H_2(t)$, one obtains the so-called Faisal-Reiss approximation
\cite{FaisalReiss}.

We want to point out that all
such series expansions are somewhat problematic since a proper convergence
of the series has not yet been established (the only known case is the Born
series in scattering theory at high energies, see e.g. \cite{GW} and the
references given there. For the one dimensional situation convergence
may be shown for integrable potentials \cite{Dav}), 
nor is it straightforward to give precise quantum
mechanical estimates of the first terms. Most statements seem to be based on
crude semi-classical estimates \cite{Mitt} or in the belief that features
which have been observed for relatively simple one dimensional models,
which are anyway put in question  \cite{Reiss}, carry
over in general \cite{Gelt1,Gelt2}.

For realistic pulses for instance with smooth adiabatic turn on and off, the
first terms will only give reasonable results when the full power is reached
but will be poor, if not completely invalid near the turn on and turn off point.

We now give a new rigorous upper bound  on the ionization probability. The proof of this
bound bypasses the problem of summing the whole perturbation series by staying
strictly with the Du Hamel formula. In what follows 
the potential $V$ will be supposed to satisfy the conditions
given in \cite{KS1} which are tailored to ensure the existence of the time
evolution operators for $H_i(t) (i=1,2,3)$ given in (2.19). In particular such potentials
$V$ are Kato small (see e.g. \cite{CFKS,RS}), i.e. there are $a < 1$
and $b < \infty$
 such that for all $\psi$ in the domain ${\cal D}(H_0)$ of $H_0= -\Delta/2 $
\begin{equation}\label{3.1}
\|V\psi\|\le a\|-\Delta\psi\|+b\|\psi\|.
\end{equation}
Also the domain ${\cal D}(H)$ of $H$ coincides with ${\cal D}(H_0)$.
We note that the potentials of atoms or molecules arising from Coulomb pair
potentials belong to this wide class.  Also all
potentials (except the $\delta$-potential) like smoothed or shielded
Coulomb potentials used in numerical computations in this context
are Kato small. Also Hamiltonians with Kato bounded
potentials (3.1) are bounded below and if the (pair) potentials decay suitably
at infinity, then there are no positive eigenvalues. Indeed, it has been shown
\cite{RS,FR} for a large class of potentials including atomic and 
molecular ones
that the eigenvalues are contained in $[inf~\sigma(H),0]$ where $\sigma(H)$
denotes the spectrum of $H$. With these specifications on V in mind
we are now in the position to state, prove and
comment on the first main result of this section.\\
\\
{\bf Upper bound 1:} 
{\it Let $\psi$ be a normalized bound state of $H=H_0+V$ with
energy $E<0$. Then for any pulse $E(t)$ with $1/2 b(\tau)^2<-E$, the ionization probability
satisfies the upper bound}
\beq \label{3.2}
I(\psi)^{1\over 2} \le \int^\tau_0 \|(V(\vec{x} -c(t) e_z) 
-V(\vec{x}))\psi\|dt  +|c(\tau)| \|p_z\psi\| \\
+ \frac{|b(\tau)|}{-E-{1\over2} b(\tau)^2} 
\|p_z\psi\|.
\eeq
\par
Note that the condition on $E(t)$ just says that the classical 
energy transfer of the
pulse is less than the classical ionization energy. 

The proof is based on a
combination of arguments used in \cite{EKS} and \cite{KS1} and goes as follows.
Since
\begin{equation}\label{3.3}
\exp -itH\psi=\exp-itE\psi \;\; ,
\end{equation}
by using the Kramers-Henneberger transformation we have
\begin{eqnarray}\label{3.4}
I(\psi)^{1\over 2} &=& \|({\bf 1}-P)U_1(\tau,0)\psi\|=\|({\bf 1}-P)T(\tau)U_3(\tau,0)\psi\|
\nonumber\\
&=& \|({\bf 1}-P)\exp-ib(\tau)z\cdot\exp ic(\tau)p_z\cdot U_3(\tau,0)\psi\|\nonumber\\
&\le&\|({\bf 1}-P)\exp-ib(\tau)z\cdot\exp ic(\tau)p_z(U_3(\tau,0)-\exp-i\tau H)\psi\|
\nonumber\\
&& + \|({\bf 1}-P)\exp-ib(\tau)z\cdot\exp ic(\tau)p_z\cdot\psi\|.
\end{eqnarray}
We start with an estimate of the first term on the r.h.s of (3.4). Obviously
it is bounded by
\begin{equation}\label{3.5}
\| (U_3(\tau,0)-\exp-i\tau H)\psi\|.
\end{equation}
We now invoke Du Hamel's formula to rewrite (3.5) as
\begin{equation}\label{3.6}
\left\|\int^\tau_0 U_3(\tau,t)[V(\vec{x}-c(t)e_z)-V(\vec{x}) 
]\exp-i(\tau-t) 
H\cdot \psi \; dt \right\|.
\end{equation}

Now 
we use the unitarity of $U_3(\tau,t)$ (besides the fact that we never
iterate Du Hamel's formula, this is the crucial step in avoiding
perturbation theory) and (3.3) to estimate (3.6) by
\begin{equation}\label{3.7}
\int^\tau_0\|(V(\vec{x} -c(t)e_z)-V(\vec{x}))\psi\|dt
\end{equation}
which is the first term on the r.h.s. of (3.2). To estimate the second term
in (\ref{3.4}),
we use the triangle inequality to obtain
\bea\label{3.8}
\|({\bf 1}-P)\exp-ib(\tau) z \cdot  \exp ic(\tau)p_z \cdot \psi\| &\le& 
\|(\exp(ic(\tau)p_z)-{\bf 1} \cdot   \psi\| \\& &
+\|({\bf 1}-P)\exp-ib(\tau)z\cdot\psi\| \nonumber
\eea
Now we use the estimate
$$
\| ( \exp iA-{\bf 1} )  \psi\|\le\| A\psi\|
$$
valid for any selfadjoint operator to estimate the first term on the r.h.s.
of (3.8) by
\begin{equation}\label{3.9}
|c(\tau)|~\| p_z\psi\|
\end{equation}
which is the second term on the r.h.s. of (3.2). It remains to estimate the
second term in (3.8). By assumption $({\bf 1}-P)H\ge 0$. Hence for any 
$\delta >~0$,$
({\bf 1}-P)(H+\delta)^{-1}$ exists and is norm bounded by $1/\delta$. Therefore
\begin{eqnarray}\label{3.10}
&&\|({\bf 1}-P)\exp-ib(\tau)z \cdot \psi\|\nonumber\\
&=&\|({\bf 1}-P)(H+\delta)^{-1} (H+\delta) \exp-ib(\tau) z\cdot\psi \|\nonumber\\
&=&\| ({\bf 1}-P)(H+\delta)^{-1} \exp-ib(\tau)z  \cdot 
 \exp~ib(\tau) z\cdot(H+\delta)\cdot
\exp-ib(\tau)z\cdot\psi\|\nonumber\\
& \le& {1\over \delta}\| \exp~ib(\tau) z\cdot(H+\delta)\cdot
\exp-ib(\tau)z\cdot\psi\|.
\end{eqnarray}
We now use the fact that
\begin{eqnarray}\label{3.11}
\exp~ib(\tau)z\cdot H\cdot\exp-ib(\tau)z&=&{1\over 2}
(-i\nabla  -b(\tau)e_z)^2+V\nonumber\\
&=& H-b(\tau)p_z+{1\over 2} b(\tau)^2.
\end{eqnarray}
Inserting this into (3.10) gives
\begin{equation}\label{3.12}
\|({\bf 1}-P)\exp-ib(\tau) z\cdot\psi\|\nonumber\\
\le {1\over\delta}\| (E-b(\tau) p_z+{1\over 2}b(\tau)^2+\delta)\psi\|
\end{equation}
Making the choice
\begin{equation}\label{3.13}
\delta=-E-{1\over 2} b(\tau)^2
\end{equation}
which by assumption on $b(\tau)$ is $>0$ and inserting into (3.12) gives the
third term in (3.2) concluding the proof of the upper bound.

We now comment on this result.

Inspection of the proof of the main result in \cite{KS1} shows that one has
the alternative
 
\noindent
{\bf Upper bound 2:}
\begin{equation}\label{3.14}
I(\psi)^{1\over 2} \le \int^\tau_0 \| (V(\vec{x} -c(t)e_z)
-V(\vec{x}))\psi\|dt
+ |c(\tau)|~\| p_z\psi\| +|b(\tau)|~\| z\psi\|
\end{equation}
which differs from (3.2) only in the last term. Typically near threshold, i.e.
for small $|E|$, both $ \|z\psi\|$ and $1/(-E-{1\over 2} b(\tau)^2)$ 
become large, whereas
$$
\| p_z\psi\| =\langle \psi,p_z^2\psi\rangle^{1\over 2} \le (2\langle\psi,
H_0\psi\rangle)^{1\over2}
$$
stays finite. Thus for $V$ being the Coulomb potential
\begin{equation}\label{3.15}
\langle\psi,H_0\psi\rangle =-E
\end{equation}
by the virial theorem (see e.g. \cite{Thirring}). 
For $s$-states $\psi_{n00}$ one can 
improve slightly since $\|p_z \psi_{n00}\|^2={2\over 3} \langle\psi_{n00},
H_0\psi_{n00}\rangle={1\over 3n^2}$. Similarly one has $\|z\psi\|^2 \le
\langle\psi,r^2\psi \rangle(={n^2\over 2} [5n^2+1-3\ell (\ell +1)]$ if 
$\psi=\psi_{n\ell m}$, see e.g. \cite{LL}). 
Again for $s$-states $\| z\psi\|^2={1\over 3}
\langle \psi,r^2\psi\rangle$, which is a slight improvement. 
Therefore (3.2) and (3.14) are essentially equivalent. 

We now discuss the first
two terms in (3.2) and (3.14). In general, for Kato bounded potentials
$V(\vec{x})(-\Delta+{\bf 1})^{-1}$ is a bounded operator. 
Also since $-\Delta$ is
translation invariant, we have
\begin{equation}\label{3.16}
\| V(\vec{x} -\vec{y}) (-\Delta +{\bf 1})^{-1} \| =\| V(\vec{x})
(-\Delta + {\bf 1})^{-1}\|.
\end{equation}
In particular for the choice of the Coulomb potential, we prove in Appendix A
that
\begin{equation}\label{3.17}
\left\|{1\over r} (-\Delta +{\bf 1})^{-1} \right\| \le 6.35.
\end{equation}
Thus for the general potentials $V$ considered, the first term in (3.2) and
(3.14) is bounded by
\bea\label{3.18}
\tau\| V(\vec{x}) (-\Delta+{\bf 1})^{-1} \|~\|(-\Delta +{\bf 1})\psi\|
&+&\tau\| V \psi \| =  \tau\| V \psi \|  \\ &+&  \tau\|V(\vec{x})
(-\Delta+{\bf 1})^{-1}\| \; \|
 (2H_0+{\bf 1})\psi\| \nonumber
\eea
which involves $E(t)$ only through its duration but not its strength. For the
eigenfunctions $\psi_{n\ell m}$ of the hydrogen atom, one has (see e.g. 
\cite{LL})
\begin{eqnarray}\label{3.19}
\|(2H_0 +{\bf 1})\psi_{n\ell m}\|^2 &=&\| ((2H+{\bf 1})-2V)\psi_{n\ell m}\|^2
\nonumber\\
&=&(2E_n+1)^2 +2(2E_n+1)\langle\psi_{n\ell m},{1\over r} \psi_{n\ell m}\rangle
 +4\langle \psi_{n\ell m},{1\over r^2}\psi_{n\ell m}\rangle\nonumber\\
&=&1-{1\over n^4}+ {4\over n^3(\ell +{1\over 2})}.
\end{eqnarray}
This quantity is $\le 8$ and behaves like $1+{\cal O} ({1\over n^3})$ 
for $n$ large
uniformly in $0\le\ell \le n-1$. Hence the right hand side of (3.18) for
bound states of the hydrogen atom is bounded by $19.4\tau$ uniformly in $n$ and
by $6.35\tau$ for all large $n$.

For the ground state $\psi_{100}$ of the hydrogen atom the first term in (3.2)
and (3.14) has a much better estimate. As shown in Appendix B
\begin{equation}\label{3.20}
\| (V(\vec{x} -\vec{y}) -V(\vec{x}))\psi_{100}\| \le 2
\end{equation}
holds for all $\vec{y}\in \Mrr^3$ such that the first term in (3.2) and (3.14)
is now bounded by $2\tau$.

By a theorem of Pfeifer \cite{P} for the survival probability $|\langle \psi,
\psi_\tau\rangle|$ of a state with $\psi_\tau =U(\tau,0)\psi$ for any 
time dependent Hamiltonian $H(t)$ one has for all small $\tau$ (see \cite{P}
for precise conditions)
\begin{equation}\label{3.21}
|\langle\psi,\psi_\tau\rangle| \ge\cos (\int^\tau_0\Delta(t)dt)
\end{equation}
where in the present context with $H(t)=H_1(t)$ (see (2.19))
\begin{eqnarray}\label{3.22}
\Delta(t)&=&|E(t)|  \cdot a_\psi \nonumber\\
a_\psi&=& (\|z\psi\|^2-\langle\psi,z\psi\rangle^2)^{1\over 2}\le\| z\psi\|.
\end{eqnarray}
This gives for the ionization probability 
\begin{eqnarray}\label{3.23}
I(\psi)&\le& 1-| \langle\psi,\psi_\tau\rangle |^2\le a_\psi^2
\left(\int_0^\tau |E(t)| dt \right)^2\nonumber\\
&\le& \left
(\int^\tau_0 |E(t)|dt \right)^2 \| z\psi\|^2.
\end{eqnarray}
This may be compared with the discussion above. (3.23) is weaker than (3.2) and
(3.14) in the sense that it does not show independence of the field strength
when $b(\tau)$ and $c(\tau)$ are small or even zero. Otherwise it is basically
equivalent to (3.2) and (3.14) or even stronger whenever the last two terms
there dominate.
We note that the rigorous bound (\ref{3.23}) may be compared with first
order perturbation theory in $E(t)$ which gives
\beq
I_{\rm pert}^{(1)} = \left( \int_0^{\tau} E(t) dt \right)^2 \| P z \psi \|^2
\leq \left| \int_0^{\tau} E(t) dt \right|^2 \| z \psi \|^2  \;\; .
\eeq

We now turn to a comparison with other approximation methods based on 
perturbative expansions used in this context.
Since all these approximations resolve around the same principle, i.e. an
expansion involving the Gordon-Volkov time evolution operator $U_{0,1}(t,t')$,
we will mainly concentrate on a recent work by Geltman \cite{Gelt1}, who
presented an explicit, partly analytical, partly numerical analysis of the
full three-dimensional hydrogen atom.
Also a nice discussion of other works may be found there. Geltman employs
the approximation method of Perelomov et al. in order to compute the
excitation and ionization rates for the $1s/2s/3s/2p/3p/3d$ states of the
hydrogen atom struck by a linearly polarized monochromatic laser pulse of the
form $E(t)=E_0 \cos \omega t$. The value for the electric field strength
in atomic units is chosen to be $E_0 =5,10,20$ and 
the frequency $\omega =1.5$ . Geltman
obtained the following general features

\begin{itemize}
\item[a)] At integer cycles, that is for $\tau=2\pi n/\omega$ the rate of
ionization becomes independent of the electric field strength. In particular
for the non-s-states it goes to zero.
\end{itemize}
This is reflected qualitatively in our results in the following way. At
integer cycles $b(\tau) =c(\tau)=0$ such that the last two terms in (3.2)
and (3.14) vanish. Also the first term is independent of the field strength, 
however, not zero and by the above discussion too large for the above 
choices of $E_0$ and $\omega$.

\begin{itemize}
\item[b)] The maxima of $b(\tau)$ and $c(\tau)$ are located at half-integer
cycles, i.e. $\tau=2\pi(n+{1\over 2})/\omega$.
\end{itemize}
For the applied pulse the bounds (3.2) and (3.14) also reproduce this
feature qualitatively but again we emphasize that these bounds hold in more
generality for all Kato potentials and all states.

Stabilization for strong, short electric fields has been a highly 
controversial issue with disagreeing results between numerous authors
on one side, for a review see \cite{Gelt3},    
as well as  on the other side, see \cite{KulS} for a review on these.

The following result
 shows absence of stabilization for sufficiently strong,
short pulses, namely when $b(\tau)$ becomes large and $\tau$ small the
ionization probability is close to 1.\\
\\
{\bf Lower bound:} {\it Let $\psi$ be a normalized bound state of 
$H=H_0 +V$ with
energy $E<0$. Then for any pulse $E(t)$ with $1/2 b(\tau)^2>-E$ the ionization probability
satisfies a lower bound of the form}
\begin{eqnarray}\label{3.24}
I(\psi)\ge 1&-&\Bigg\{ \int^\tau_0\| (V(\vec{x} 
-c(t)e_z)-V(\vec{x}))\psi\|dt\\
&&+ \frac{1}{E+{1\over 2} b(\tau)^2} \| 
(V(\vec{x} -c(\tau)e_z)-V(\vec{x}))\psi\|
+\frac{|b(\tau)|}{E+{1\over 2} b(\tau)^2} \|p_z\psi\|\Bigg\}^2.
\nonumber
\end{eqnarray}
Note that now the condition on $E(t)$ is that the classical energy transfer
of the pulse is larger than the classical ionization energy.
Recall that by our previous discussion the norms appearing in the first two
terms in the bracket may be estimated independently of the field strength, such
that (\ref{3.24}) gives a bound which involves $E(t)$ only 
through $\tau$ and $b(\tau)$.

We turn to a proof. In order to obtain a lower bound on
$$
I(\psi) =\| ({\bf 1}-P) U_1(\tau,0) \psi\|^2 = 1-\|PU_1(\tau,0)\psi\|^2
$$
it suffices to obtain an upper bound on $\|PU_1(\tau,0)\psi\|$. First we write
\begin{eqnarray}\label{3.25}
\|PU_1(\tau,0)\psi\| &=&\|PT(\tau)U_3(\tau,0)\psi\| \nonumber\\
&=&\|P\exp-ib(\tau)z\cdot\exp ic(\tau)p_z\cdot U_3(\tau,0)\psi\|\nonumber\\
&\le& \|P\exp-ib(\tau)z\cdot\exp ic(\tau)p_z\cdot(U_3(\tau,0)-\exp-i\tau H)\psi\|\nonumber\\
&&+\|P\exp-ib(\tau)z\cdot\exp ic(\tau)p_z\cdot\psi\|.
\end{eqnarray}
The first term on the r.h.s. is estimated by (3.5) yielding the first term 
in the bracket in (\ref{3.24}).

The second term  in (\ref{3.25})
is treated as follows. By assumption $PH\le 0$. Let $\delta >0$
be arbitrary. Then $P(H-\delta)^{-1}$ is a well defined operator with 
operator norm $\le 1/\delta$. Hence 
\begin{eqnarray}\label{3.26}
&&\|P\exp-ib(\tau)z \cdot \exp ic(\tau)p_z\cdot\psi\|\nonumber\\
&=&\| P(H-\delta)^{-1} (H-\delta)\exp-ib(\tau)z\cdot\exp ic(\tau)p_z\cdot\psi\|\nonumber\\
&\le& {1\over \delta}\| (H-\delta) 
\exp-ib(\tau)z\cdot\exp ic(\tau)p_z\cdot\psi\|.
\end{eqnarray}
In analogy to (3.11) we now use the relation 
\begin{eqnarray}\label{3.27}
\lefteqn{\exp- ic(\tau)p_z\cdot\exp~ib(\tau)z\cdot H\cdot
\exp-ib(\tau)z\cdot\exp ic(\tau)
p_z}\nonumber\\
&& ={1\over 2} (-i\nabla -b(\tau)e_z)^2 +V(\vec{x}
-c(\tau) e_z)\nonumber\\
&& =H-b(\tau)p_z+{1\over 2} b(\tau)^2+V(\vec{x} -c(\tau)e_z)-V(\vec{x}).
\end{eqnarray}
Inserting this into (\ref{3.26}) we obtain
\begin{eqnarray}\label{3.28}
\!\!\!\!\!\!\!\!\!\!\!\! \| P\exp-ib(\tau)z \cdot 
\exp ic(\tau)p_z\psi\| &\le& 
{1\over\delta}\|(V(\vec{x}-c(\tau)e_z)-V(\vec{x}) \cdot \psi\| 
\nonumber\\ && +{1\over\delta}\|
(E-b(\tau)p_z +{1\over 2} b(\tau)^2-\delta)\psi\|
\end{eqnarray}
We now  make the choice
\begin{equation}\label{3.29}
\delta =E+{1\over 2} b(\tau)^2
\end{equation}
which by assumption on $b(\tau)$ is $>0$ and when inserted into 
(\ref{3.28}) immediately yields
the remaining two terms in the bracket of (\ref{3.24}), 
thus concluding the proof of the lower bound.

We note that these two theorems are compatible with the result in \cite{EKS}
on the Stark kick, i.e. $E(t)=F_0\delta(t)$. There it was shown that for
fixed ionization probability of any bound state $\psi_{n\ell m}$ of the
hydrogen atom $F_0$ scales like $1/n$, as predicted by Rheinhold et al.
\cite{Rhein}.

We now return to a comparison between our results and those obtained
by employing approximation methods based on perturbative expansions
and we will include the lower bound in the discussion. 
We stress once more the
point that the lower bound definitely excludes the possibility of stabilization
of the bound states for increasing electric field strength, when the
applied pulse is short in duration, since (\ref{3.7}), despite the
dependence on $c(t)$, may be estimated by a constant, say C, independent
of the electric field. Hence
\beq
\lim_{|E(t)| \rightarrow \infty} I(\psi) \;\geq \; 1 - \tau^2 \; C \;\; .
\eeq
This shows clearly that 
the electric field has no stabilizing effect and we are therefore
in disagreement with \cite{Kul,Horbatsch,Pont}. The lower bound also
reproduces the result obtained through an expansion around the 
Gordon-Volkov solution, namely for  monochromatic linearly polarized 
laser pulses at integer cycles, i.e. $b(\tau) = c(\tau) =0 $, the
ionization probability becomes independent of the electric field
strength. We may qualitatively relate the term proportional to $\tau^2$ 
to a term also observed in  perturbative expansion methods and which is
interpreted as the spreading of the wave.

To illustrate our results further, we consider now the concrete
example of the hydrogen atom. For the $\psi_{100}$ state we
obtain as our best estimate
\bea
I(\psi_{100}) &\leq&\left( 2\tau +  |b(\tau)|
+ \frac{1}{\sqrt{3}} |c(\tau)| \right)^2 \\
I(\psi_{100}) &\geq&1 - \left( 2\tau + \frac{4}{b(\tau)^2 -1} 
+ \frac{2}{\sqrt{3}} \frac{|b(\tau)|}{b(\tau)^2 -1} \right)^2\; .
\eea
Taking the pulse to be of the form $E_z(t) = E_0 \cos \omega t$ 
for $0 \leq t \leq \tau$ and zero otherwise,   we
have
\beq
|b(\tau)| = \frac{E_0}{\omega} |\sin \omega \tau  | \qquad
\hbox{and} \qquad
|c(\tau)| = \frac{2E_0}{\omega^2} \sin^2 \left( 
\frac{\omega \tau}{2}\right)\; .
\eeq
\begin{figure}[h]
   \begin{center}
\setlength{\unitlength}{0.240900pt}
\ifx\plotpoint\undefined\newsavebox{\plotpoint}\fi
\sbox{\plotpoint}{\rule[-0.200pt]{0.400pt}{0.400pt}}%
\begin{picture}(1500,900)(0,0)
\font\gnuplot=cmr10 at 10pt
\gnuplot
\sbox{\plotpoint}{\rule[-0.200pt]{0.400pt}{0.400pt}}%
\put(220.0,113.0){\rule[-0.200pt]{292.934pt}{0.400pt}}
\put(220.0,113.0){\rule[-0.200pt]{0.400pt}{184.048pt}}
\put(220.0,113.0){\rule[-0.200pt]{4.818pt}{0.400pt}}
\put(198,113){\makebox(0,0)[r]{0}}
\put(1416.0,113.0){\rule[-0.200pt]{4.818pt}{0.400pt}}
\put(220.0,266.0){\rule[-0.200pt]{4.818pt}{0.400pt}}
\put(198,266){\makebox(0,0)[r]{0.2}}
\put(1416.0,266.0){\rule[-0.200pt]{4.818pt}{0.400pt}}
\put(220.0,419.0){\rule[-0.200pt]{4.818pt}{0.400pt}}
\put(198,419){\makebox(0,0)[r]{0.4}}
\put(1416.0,419.0){\rule[-0.200pt]{4.818pt}{0.400pt}}
\put(220.0,571.0){\rule[-0.200pt]{4.818pt}{0.400pt}}
\put(198,571){\makebox(0,0)[r]{0.6}}
\put(1416.0,571.0){\rule[-0.200pt]{4.818pt}{0.400pt}}
\put(220.0,724.0){\rule[-0.200pt]{4.818pt}{0.400pt}}
\put(198,724){\makebox(0,0)[r]{0.8}}
\put(1416.0,724.0){\rule[-0.200pt]{4.818pt}{0.400pt}}
\put(220.0,877.0){\rule[-0.200pt]{4.818pt}{0.400pt}}
\put(198,877){\makebox(0,0)[r]{1}}
\put(1416.0,877.0){\rule[-0.200pt]{4.818pt}{0.400pt}}
\put(220.0,113.0){\rule[-0.200pt]{0.400pt}{4.818pt}}
\put(220,68){\makebox(0,0){0}}
\put(220.0,857.0){\rule[-0.200pt]{0.400pt}{4.818pt}}
\put(365.0,113.0){\rule[-0.200pt]{0.400pt}{4.818pt}}
\put(365,68){\makebox(0,0){0.5}}
\put(365.0,857.0){\rule[-0.200pt]{0.400pt}{4.818pt}}
\put(510.0,113.0){\rule[-0.200pt]{0.400pt}{4.818pt}}
\put(510,68){\makebox(0,0){1}}
\put(510.0,857.0){\rule[-0.200pt]{0.400pt}{4.818pt}}
\put(655.0,113.0){\rule[-0.200pt]{0.400pt}{4.818pt}}
\put(655,68){\makebox(0,0){1.5}}
\put(655.0,857.0){\rule[-0.200pt]{0.400pt}{4.818pt}}
\put(801.0,113.0){\rule[-0.200pt]{0.400pt}{4.818pt}}
\put(801,68){\makebox(0,0){2}}
\put(801.0,857.0){\rule[-0.200pt]{0.400pt}{4.818pt}}
\put(946.0,113.0){\rule[-0.200pt]{0.400pt}{4.818pt}}
\put(946,68){\makebox(0,0){2.5}}
\put(946.0,857.0){\rule[-0.200pt]{0.400pt}{4.818pt}}
\put(1091.0,113.0){\rule[-0.200pt]{0.400pt}{4.818pt}}
\put(1091,68){\makebox(0,0){3}}
\put(1091.0,857.0){\rule[-0.200pt]{0.400pt}{4.818pt}}
\put(1236.0,113.0){\rule[-0.200pt]{0.400pt}{4.818pt}}
\put(1236,68){\makebox(0,0){3.5}}
\put(1236.0,857.0){\rule[-0.200pt]{0.400pt}{4.818pt}}
\put(1381.0,113.0){\rule[-0.200pt]{0.400pt}{4.818pt}}
\put(1381,68){\makebox(0,0){4}}
\put(1381.0,857.0){\rule[-0.200pt]{0.400pt}{4.818pt}}
\put(220.0,113.0){\rule[-0.200pt]{292.934pt}{0.400pt}}
\put(1436.0,113.0){\rule[-0.200pt]{0.400pt}{184.048pt}}
\put(220.0,877.0){\rule[-0.200pt]{292.934pt}{0.400pt}}
\put(45,495){\makebox(0,0){I}}
\put(828,23){\makebox(0,0){time}}
\put(220.0,113.0){\rule[-0.200pt]{0.400pt}{184.048pt}}
\put(220,113){\usebox{\plotpoint}}
\multiput(220.58,113.00)(0.492,1.487){21}{\rule{0.119pt}{1.267pt}}
\multiput(219.17,113.00)(12.000,32.371){2}{\rule{0.400pt}{0.633pt}}
\multiput(232.58,148.00)(0.493,4.263){23}{\rule{0.119pt}{3.423pt}}
\multiput(231.17,148.00)(13.000,100.895){2}{\rule{0.400pt}{1.712pt}}
\multiput(245.58,256.00)(0.492,7.906){21}{\rule{0.119pt}{6.233pt}}
\multiput(244.17,256.00)(12.000,171.062){2}{\rule{0.400pt}{3.117pt}}
\multiput(257.58,440.00)(0.492,11.267){21}{\rule{0.119pt}{8.833pt}}
\multiput(256.17,440.00)(12.000,243.666){2}{\rule{0.400pt}{4.417pt}}
\multiput(269.59,702.00)(0.482,15.769){9}{\rule{0.116pt}{11.767pt}}
\multiput(268.17,702.00)(6.000,150.578){2}{\rule{0.400pt}{5.883pt}}
\multiput(1381.59,828.16)(0.482,-15.769){9}{\rule{0.116pt}{11.767pt}}
\multiput(1380.17,852.58)(6.000,-150.578){2}{\rule{0.400pt}{5.883pt}}
\multiput(1387.58,665.33)(0.492,-11.267){21}{\rule{0.119pt}{8.833pt}}
\multiput(1386.17,683.67)(12.000,-243.666){2}{\rule{0.400pt}{4.417pt}}
\multiput(1399.58,414.12)(0.492,-7.906){21}{\rule{0.119pt}{6.233pt}}
\multiput(1398.17,427.06)(12.000,-171.062){2}{\rule{0.400pt}{3.117pt}}
\multiput(1411.58,241.79)(0.493,-4.263){23}{\rule{0.119pt}{3.423pt}}
\multiput(1410.17,248.90)(13.000,-100.895){2}{\rule{0.400pt}{1.712pt}}
\multiput(1424.58,142.74)(0.492,-1.487){21}{\rule{0.119pt}{1.267pt}}
\multiput(1423.17,145.37)(12.000,-32.371){2}{\rule{0.400pt}{0.633pt}}
\put(220,113){\usebox{\plotpoint}}
\multiput(220,113)(1.773,20.680){7}{\usebox{\plotpoint}}
\multiput(232,253)(0.626,20.746){21}{\usebox{\plotpoint}}
\multiput(245,684)(0.323,20.753){9}{\usebox{\plotpoint}}
\put(248,877){\usebox{\plotpoint}}
\multiput(1408,877)(0.323,-20.753){10}{\usebox{\plotpoint}}
\multiput(1411,684)(0.626,-20.746){21}{\usebox{\plotpoint}}
\multiput(1424,253)(1.773,-20.680){6}{\usebox{\plotpoint}}
\put(1436,113){\usebox{\plotpoint}}
\sbox{\plotpoint}{\rule[-0.400pt]{0.800pt}{0.800pt}}%
\put(220,113){\usebox{\plotpoint}}
\multiput(221.41,113.00)(0.511,25.262){17}{\rule{0.123pt}{37.533pt}}
\multiput(218.34,113.00)(12.000,482.098){2}{\rule{0.800pt}{18.767pt}}
\put(231.34,673){\rule{0.800pt}{49.144pt}}
\multiput(230.34,673.00)(2.000,102.000){2}{\rule{0.800pt}{24.572pt}}
\put(1421.34,673){\rule{0.800pt}{49.144pt}}
\multiput(1420.34,775.00)(2.000,-102.000){2}{\rule{0.800pt}{24.572pt}}
\multiput(1425.41,517.20)(0.511,-25.262){17}{\rule{0.123pt}{37.533pt}}
\multiput(1422.34,595.10)(12.000,-482.098){2}{\rule{0.800pt}{18.767pt}}
\sbox{\plotpoint}{\rule[-0.500pt]{1.000pt}{1.000pt}}%
\multiput(423,113)(2.628,20.588){3}{\usebox{\plotpoint}}
\multiput(429,160)(3.196,20.508){4}{\usebox{\plotpoint}}
\multiput(441,237)(4.276,20.310){2}{\usebox{\plotpoint}}
\multiput(453,294)(6.006,19.867){3}{\usebox{\plotpoint}}
\put(473.23,356.27){\usebox{\plotpoint}}
\put(481.31,375.35){\usebox{\plotpoint}}
\put(491.29,393.49){\usebox{\plotpoint}}
\put(505.40,408.60){\usebox{\plotpoint}}
\put(524.18,415.77){\usebox{\plotpoint}}
\multiput(527,416)(19.690,-6.563){0}{\usebox{\plotpoint}}
\put(542.88,408.42){\usebox{\plotpoint}}
\put(556.63,393.06){\usebox{\plotpoint}}
\multiput(564,382)(8.176,-19.077){2}{\usebox{\plotpoint}}
\put(581.71,336.39){\usebox{\plotpoint}}
\multiput(588,317)(5.223,-20.088){3}{\usebox{\plotpoint}}
\multiput(601,267)(3.713,-20.421){3}{\usebox{\plotpoint}}
\multiput(613,201)(2.804,-20.565){4}{\usebox{\plotpoint}}
\put(625,113){\usebox{\plotpoint}}
\multiput(1031,113)(2.804,20.565){5}{\usebox{\plotpoint}}
\multiput(1043,201)(3.713,20.421){3}{\usebox{\plotpoint}}
\multiput(1055,267)(5.223,20.088){3}{\usebox{\plotpoint}}
\put(1074.40,336.74){\usebox{\plotpoint}}
\multiput(1080,354)(8.176,19.077){2}{\usebox{\plotpoint}}
\put(1099.58,393.36){\usebox{\plotpoint}}
\put(1113.39,408.67){\usebox{\plotpoint}}
\multiput(1117,412)(19.690,6.563){0}{\usebox{\plotpoint}}
\put(1132.18,415.73){\usebox{\plotpoint}}
\put(1150.91,408.39){\usebox{\plotpoint}}
\put(1164.95,393.22){\usebox{\plotpoint}}
\put(1174.86,375.02){\usebox{\plotpoint}}
\put(1182.90,355.93){\usebox{\plotpoint}}
\multiput(1190,337)(6.006,-19.867){3}{\usebox{\plotpoint}}
\multiput(1203,294)(4.276,-20.310){2}{\usebox{\plotpoint}}
\multiput(1215,237)(3.196,-20.508){4}{\usebox{\plotpoint}}
\multiput(1227,160)(2.628,-20.588){2}{\usebox{\plotpoint}}
\put(1233,113){\usebox{\plotpoint}}
\sbox{\plotpoint}{\rule[-0.600pt]{1.200pt}{1.200pt}}%
\multiput(309.24,113.00)(0.502,12.670){12}{\rule{0.121pt}{28.336pt}}
\multiput(304.51,113.00)(11.000,198.186){2}{\rule{1.200pt}{14.168pt}}
\multiput(320.24,370.00)(0.501,6.209){16}{\rule{0.121pt}{14.423pt}}
\multiput(315.51,370.00)(13.000,123.064){2}{\rule{1.200pt}{7.212pt}}
\multiput(333.24,523.00)(0.501,4.043){14}{\rule{0.121pt}{9.500pt}}
\multiput(328.51,523.00)(12.000,72.282){2}{\rule{1.200pt}{4.750pt}}
\multiput(345.24,615.00)(0.501,2.559){14}{\rule{0.121pt}{6.200pt}}
\multiput(340.51,615.00)(12.000,46.132){2}{\rule{1.200pt}{3.100pt}}
\multiput(357.24,674.00)(0.501,1.704){14}{\rule{0.121pt}{4.300pt}}
\multiput(352.51,674.00)(12.000,31.075){2}{\rule{1.200pt}{2.150pt}}
\multiput(369.24,714.00)(0.501,1.068){16}{\rule{0.121pt}{2.885pt}}
\multiput(364.51,714.00)(13.000,22.013){2}{\rule{1.200pt}{1.442pt}}
\multiput(382.24,742.00)(0.501,0.849){14}{\rule{0.121pt}{2.400pt}}
\multiput(377.51,742.00)(12.000,16.019){2}{\rule{1.200pt}{1.200pt}}
\multiput(394.24,763.00)(0.501,0.579){14}{\rule{0.121pt}{1.800pt}}
\multiput(389.51,763.00)(12.000,11.264){2}{\rule{1.200pt}{0.900pt}}
\multiput(404.00,780.24)(0.533,0.502){12}{\rule{1.718pt}{0.121pt}}
\multiput(404.00,775.51)(9.434,11.000){2}{\rule{0.859pt}{1.200pt}}
\multiput(417.00,791.24)(0.588,0.502){8}{\rule{1.900pt}{0.121pt}}
\multiput(417.00,786.51)(8.056,9.000){2}{\rule{0.950pt}{1.200pt}}
\multiput(429.00,800.24)(0.738,0.505){4}{\rule{2.357pt}{0.122pt}}
\multiput(429.00,795.51)(7.108,7.000){2}{\rule{1.179pt}{1.200pt}}
\put(441,805.01){\rule{2.891pt}{1.200pt}}
\multiput(441.00,802.51)(6.000,5.000){2}{\rule{1.445pt}{1.200pt}}
\put(453,809.51){\rule{3.132pt}{1.200pt}}
\multiput(453.00,807.51)(6.500,4.000){2}{\rule{1.566pt}{1.200pt}}
\put(466,813.01){\rule{2.891pt}{1.200pt}}
\multiput(466.00,811.51)(6.000,3.000){2}{\rule{1.445pt}{1.200pt}}
\put(478,815.51){\rule{2.891pt}{1.200pt}}
\multiput(478.00,814.51)(6.000,2.000){2}{\rule{1.445pt}{1.200pt}}
\put(490,817.51){\rule{3.132pt}{1.200pt}}
\multiput(490.00,816.51)(6.500,2.000){2}{\rule{1.566pt}{1.200pt}}
\put(515,819.01){\rule{2.891pt}{1.200pt}}
\multiput(515.00,818.51)(6.000,1.000){2}{\rule{1.445pt}{1.200pt}}
\put(527,819.01){\rule{2.891pt}{1.200pt}}
\multiput(527.00,819.51)(6.000,-1.000){2}{\rule{1.445pt}{1.200pt}}
\put(539,818.01){\rule{3.132pt}{1.200pt}}
\multiput(539.00,818.51)(6.500,-1.000){2}{\rule{1.566pt}{1.200pt}}
\put(552,816.51){\rule{2.891pt}{1.200pt}}
\multiput(552.00,817.51)(6.000,-2.000){2}{\rule{1.445pt}{1.200pt}}
\put(564,814.51){\rule{2.891pt}{1.200pt}}
\multiput(564.00,815.51)(6.000,-2.000){2}{\rule{1.445pt}{1.200pt}}
\put(576,811.51){\rule{2.891pt}{1.200pt}}
\multiput(576.00,813.51)(6.000,-4.000){2}{\rule{1.445pt}{1.200pt}}
\put(588,807.51){\rule{3.132pt}{1.200pt}}
\multiput(588.00,809.51)(6.500,-4.000){2}{\rule{1.566pt}{1.200pt}}
\multiput(601.00,805.25)(0.792,-0.509){2}{\rule{2.700pt}{0.123pt}}
\multiput(601.00,805.51)(6.396,-6.000){2}{\rule{1.350pt}{1.200pt}}
\multiput(613.00,799.26)(0.657,-0.503){6}{\rule{2.100pt}{0.121pt}}
\multiput(613.00,799.51)(7.641,-8.000){2}{\rule{1.050pt}{1.200pt}}
\multiput(625.00,791.26)(0.587,-0.502){10}{\rule{1.860pt}{0.121pt}}
\multiput(625.00,791.51)(9.139,-10.000){2}{\rule{0.930pt}{1.200pt}}
\multiput(640.24,777.36)(0.501,-0.489){14}{\rule{0.121pt}{1.600pt}}
\multiput(635.51,780.68)(12.000,-9.679){2}{\rule{1.200pt}{0.800pt}}
\multiput(652.24,762.28)(0.501,-0.714){14}{\rule{0.121pt}{2.100pt}}
\multiput(647.51,766.64)(12.000,-13.641){2}{\rule{1.200pt}{1.050pt}}
\multiput(664.24,741.79)(0.501,-0.984){14}{\rule{0.121pt}{2.700pt}}
\multiput(659.51,747.40)(12.000,-18.396){2}{\rule{1.200pt}{1.350pt}}
\multiput(676.24,715.11)(0.501,-1.274){16}{\rule{0.121pt}{3.346pt}}
\multiput(671.51,722.05)(13.000,-26.055){2}{\rule{1.200pt}{1.673pt}}
\multiput(689.24,674.83)(0.501,-2.064){14}{\rule{0.121pt}{5.100pt}}
\multiput(684.51,685.41)(12.000,-37.415){2}{\rule{1.200pt}{2.550pt}}
\multiput(701.24,616.04)(0.501,-3.234){14}{\rule{0.121pt}{7.700pt}}
\multiput(696.51,632.02)(12.000,-58.018){2}{\rule{1.200pt}{3.850pt}}
\multiput(713.24,527.92)(0.501,-4.728){16}{\rule{0.121pt}{11.100pt}}
\multiput(708.51,550.96)(13.000,-93.961){2}{\rule{1.200pt}{5.550pt}}
\multiput(726.24,371.07)(0.501,-9.082){14}{\rule{0.121pt}{20.700pt}}
\multiput(721.51,414.04)(12.000,-161.036){2}{\rule{1.200pt}{10.350pt}}
\put(735.51,113){\rule{1.200pt}{33.726pt}}
\multiput(733.51,183.00)(4.000,-70.000){2}{\rule{1.200pt}{16.863pt}}
\put(915.51,113){\rule{1.200pt}{33.726pt}}
\multiput(913.51,113.00)(4.000,70.000){2}{\rule{1.200pt}{16.863pt}}
\multiput(922.24,253.00)(0.501,9.082){14}{\rule{0.121pt}{20.700pt}}
\multiput(917.51,253.00)(12.000,161.036){2}{\rule{1.200pt}{10.350pt}}
\multiput(934.24,457.00)(0.501,4.728){16}{\rule{0.121pt}{11.100pt}}
\multiput(929.51,457.00)(13.000,93.961){2}{\rule{1.200pt}{5.550pt}}
\multiput(947.24,574.00)(0.501,3.234){14}{\rule{0.121pt}{7.700pt}}
\multiput(942.51,574.00)(12.000,58.018){2}{\rule{1.200pt}{3.850pt}}
\multiput(959.24,648.00)(0.501,2.064){14}{\rule{0.121pt}{5.100pt}}
\multiput(954.51,648.00)(12.000,37.415){2}{\rule{1.200pt}{2.550pt}}
\multiput(971.24,696.00)(0.501,1.274){16}{\rule{0.121pt}{3.346pt}}
\multiput(966.51,696.00)(13.000,26.055){2}{\rule{1.200pt}{1.673pt}}
\multiput(984.24,729.00)(0.501,0.984){14}{\rule{0.121pt}{2.700pt}}
\multiput(979.51,729.00)(12.000,18.396){2}{\rule{1.200pt}{1.350pt}}
\multiput(996.24,753.00)(0.501,0.714){14}{\rule{0.121pt}{2.100pt}}
\multiput(991.51,753.00)(12.000,13.641){2}{\rule{1.200pt}{1.050pt}}
\multiput(1008.24,771.00)(0.501,0.489){14}{\rule{0.121pt}{1.600pt}}
\multiput(1003.51,771.00)(12.000,9.679){2}{\rule{1.200pt}{0.800pt}}
\multiput(1018.00,786.24)(0.587,0.502){10}{\rule{1.860pt}{0.121pt}}
\multiput(1018.00,781.51)(9.139,10.000){2}{\rule{0.930pt}{1.200pt}}
\multiput(1031.00,796.24)(0.657,0.503){6}{\rule{2.100pt}{0.121pt}}
\multiput(1031.00,791.51)(7.641,8.000){2}{\rule{1.050pt}{1.200pt}}
\multiput(1043.00,804.24)(0.792,0.509){2}{\rule{2.700pt}{0.123pt}}
\multiput(1043.00,799.51)(6.396,6.000){2}{\rule{1.350pt}{1.200pt}}
\put(1055,807.51){\rule{3.132pt}{1.200pt}}
\multiput(1055.00,805.51)(6.500,4.000){2}{\rule{1.566pt}{1.200pt}}
\put(1068,811.51){\rule{2.891pt}{1.200pt}}
\multiput(1068.00,809.51)(6.000,4.000){2}{\rule{1.445pt}{1.200pt}}
\put(1080,814.51){\rule{2.891pt}{1.200pt}}
\multiput(1080.00,813.51)(6.000,2.000){2}{\rule{1.445pt}{1.200pt}}
\put(1092,816.51){\rule{2.891pt}{1.200pt}}
\multiput(1092.00,815.51)(6.000,2.000){2}{\rule{1.445pt}{1.200pt}}
\put(1104,818.01){\rule{3.132pt}{1.200pt}}
\multiput(1104.00,817.51)(6.500,1.000){2}{\rule{1.566pt}{1.200pt}}
\put(1117,819.01){\rule{2.891pt}{1.200pt}}
\multiput(1117.00,818.51)(6.000,1.000){2}{\rule{1.445pt}{1.200pt}}
\put(1129,819.01){\rule{2.891pt}{1.200pt}}
\multiput(1129.00,819.51)(6.000,-1.000){2}{\rule{1.445pt}{1.200pt}}
\put(503.0,821.0){\rule[-0.600pt]{2.891pt}{1.200pt}}
\put(1153,817.51){\rule{3.132pt}{1.200pt}}
\multiput(1153.00,818.51)(6.500,-2.000){2}{\rule{1.566pt}{1.200pt}}
\put(1166,815.51){\rule{2.891pt}{1.200pt}}
\multiput(1166.00,816.51)(6.000,-2.000){2}{\rule{1.445pt}{1.200pt}}
\put(1178,813.01){\rule{2.891pt}{1.200pt}}
\multiput(1178.00,814.51)(6.000,-3.000){2}{\rule{1.445pt}{1.200pt}}
\put(1190,809.51){\rule{3.132pt}{1.200pt}}
\multiput(1190.00,811.51)(6.500,-4.000){2}{\rule{1.566pt}{1.200pt}}
\put(1203,805.01){\rule{2.891pt}{1.200pt}}
\multiput(1203.00,807.51)(6.000,-5.000){2}{\rule{1.445pt}{1.200pt}}
\multiput(1215.00,802.26)(0.738,-0.505){4}{\rule{2.357pt}{0.122pt}}
\multiput(1215.00,802.51)(7.108,-7.000){2}{\rule{1.179pt}{1.200pt}}
\multiput(1227.00,795.26)(0.588,-0.502){8}{\rule{1.900pt}{0.121pt}}
\multiput(1227.00,795.51)(8.056,-9.000){2}{\rule{0.950pt}{1.200pt}}
\multiput(1239.00,786.26)(0.533,-0.502){12}{\rule{1.718pt}{0.121pt}}
\multiput(1239.00,786.51)(9.434,-11.000){2}{\rule{0.859pt}{1.200pt}}
\multiput(1254.24,770.53)(0.501,-0.579){14}{\rule{0.121pt}{1.800pt}}
\multiput(1249.51,774.26)(12.000,-11.264){2}{\rule{1.200pt}{0.900pt}}
\multiput(1266.24,753.04)(0.501,-0.849){14}{\rule{0.121pt}{2.400pt}}
\multiput(1261.51,758.02)(12.000,-16.019){2}{\rule{1.200pt}{1.200pt}}
\multiput(1278.24,730.03)(0.501,-1.068){16}{\rule{0.121pt}{2.885pt}}
\multiput(1273.51,736.01)(13.000,-22.013){2}{\rule{1.200pt}{1.442pt}}
\multiput(1291.24,696.15)(0.501,-1.704){14}{\rule{0.121pt}{4.300pt}}
\multiput(1286.51,705.08)(12.000,-31.075){2}{\rule{1.200pt}{2.150pt}}
\multiput(1303.24,648.26)(0.501,-2.559){14}{\rule{0.121pt}{6.200pt}}
\multiput(1298.51,661.13)(12.000,-46.132){2}{\rule{1.200pt}{3.100pt}}
\multiput(1315.24,575.56)(0.501,-4.043){14}{\rule{0.121pt}{9.500pt}}
\multiput(1310.51,595.28)(12.000,-72.282){2}{\rule{1.200pt}{4.750pt}}
\multiput(1327.24,463.13)(0.501,-6.209){16}{\rule{0.121pt}{14.423pt}}
\multiput(1322.51,493.06)(13.000,-123.064){2}{\rule{1.200pt}{7.212pt}}
\multiput(1340.24,252.37)(0.502,-12.670){12}{\rule{0.121pt}{28.336pt}}
\multiput(1335.51,311.19)(11.000,-198.186){2}{\rule{1.200pt}{14.168pt}}
\put(1141.0,821.0){\rule[-0.600pt]{2.891pt}{1.200pt}}
\sbox{\plotpoint}{\rule[-0.500pt]{1.000pt}{1.000pt}}%
\multiput(264,113)(0.667,41.506){8}{\usebox{\plotpoint}}
\multiput(269,424)(2.163,41.455){6}{\usebox{\plotpoint}}
\multiput(281,654)(5.747,41.111){2}{\usebox{\plotpoint}}
\put(301.45,774.94){\usebox{\plotpoint}}
\put(316.32,813.51){\usebox{\plotpoint}}
\multiput(318,817)(27.187,31.369){0}{\usebox{\plotpoint}}
\multiput(331,832)(31.890,26.575){0}{\usebox{\plotpoint}}
\put(344.87,843.09){\usebox{\plotpoint}}
\multiput(355,849)(38.318,15.966){0}{\usebox{\plotpoint}}
\multiput(367,854)(40.448,9.334){0}{\usebox{\plotpoint}}
\put(383.34,857.84){\usebox{\plotpoint}}
\multiput(392,860)(40.946,6.824){0}{\usebox{\plotpoint}}
\multiput(404,862)(41.389,3.184){0}{\usebox{\plotpoint}}
\put(424.36,863.61){\usebox{\plotpoint}}
\multiput(429,864)(41.368,3.447){0}{\usebox{\plotpoint}}
\multiput(441,865)(41.368,3.447){0}{\usebox{\plotpoint}}
\put(465.73,866.98){\usebox{\plotpoint}}
\multiput(466,867)(41.511,0.000){0}{\usebox{\plotpoint}}
\multiput(478,867)(41.511,0.000){0}{\usebox{\plotpoint}}
\multiput(490,867)(41.389,3.184){0}{\usebox{\plotpoint}}
\put(507.20,868.00){\usebox{\plotpoint}}
\multiput(515,868)(41.511,0.000){0}{\usebox{\plotpoint}}
\multiput(527,868)(41.511,0.000){0}{\usebox{\plotpoint}}
\put(548.71,868.00){\usebox{\plotpoint}}
\multiput(552,868)(41.368,-3.447){0}{\usebox{\plotpoint}}
\multiput(564,867)(41.511,0.000){0}{\usebox{\plotpoint}}
\multiput(576,867)(41.368,-3.447){0}{\usebox{\plotpoint}}
\put(590.14,866.00){\usebox{\plotpoint}}
\multiput(601,866)(41.368,-3.447){0}{\usebox{\plotpoint}}
\multiput(613,865)(41.368,-3.447){0}{\usebox{\plotpoint}}
\put(631.55,863.50){\usebox{\plotpoint}}
\multiput(638,863)(40.946,-6.824){0}{\usebox{\plotpoint}}
\multiput(650,861)(40.946,-6.824){0}{\usebox{\plotpoint}}
\put(672.39,856.40){\usebox{\plotpoint}}
\multiput(674,856)(38.744,-14.902){0}{\usebox{\plotpoint}}
\multiput(687,851)(38.318,-15.966){0}{\usebox{\plotpoint}}
\put(709.75,838.83){\usebox{\plotpoint}}
\multiput(711,838)(30.502,-28.156){0}{\usebox{\plotpoint}}
\put(735.48,806.86){\usebox{\plotpoint}}
\multiput(736,806)(14.186,-39.012){0}{\usebox{\plotpoint}}
\multiput(748,773)(7.767,-40.778){2}{\usebox{\plotpoint}}
\multiput(760,710)(3.784,-41.338){3}{\usebox{\plotpoint}}
\multiput(773,568)(1.220,-41.493){10}{\usebox{\plotpoint}}
\put(785.00,146.84){\usebox{\plotpoint}}
\put(785,113){\usebox{\plotpoint}}
\multiput(871,113)(0.000,41.511){2}{\usebox{\plotpoint}}
\multiput(871,160)(1.220,41.493){9}{\usebox{\plotpoint}}
\multiput(883,568)(3.784,41.338){4}{\usebox{\plotpoint}}
\put(900.66,734.46){\usebox{\plotpoint}}
\put(908.78,775.14){\usebox{\plotpoint}}
\put(924.46,813.44){\usebox{\plotpoint}}
\multiput(932,826)(30.502,28.156){0}{\usebox{\plotpoint}}
\put(952.63,843.09){\usebox{\plotpoint}}
\multiput(957,846)(38.318,15.966){0}{\usebox{\plotpoint}}
\multiput(969,851)(38.744,14.902){0}{\usebox{\plotpoint}}
\put(991.05,858.26){\usebox{\plotpoint}}
\multiput(994,859)(40.946,6.824){0}{\usebox{\plotpoint}}
\multiput(1006,861)(40.946,6.824){0}{\usebox{\plotpoint}}
\multiput(1018,863)(41.389,3.184){0}{\usebox{\plotpoint}}
\put(1032.10,864.09){\usebox{\plotpoint}}
\multiput(1043,865)(41.368,3.447){0}{\usebox{\plotpoint}}
\multiput(1055,866)(41.511,0.000){0}{\usebox{\plotpoint}}
\put(1073.51,866.46){\usebox{\plotpoint}}
\multiput(1080,867)(41.511,0.000){0}{\usebox{\plotpoint}}
\multiput(1092,867)(41.368,3.447){0}{\usebox{\plotpoint}}
\put(1114.96,868.00){\usebox{\plotpoint}}
\multiput(1117,868)(41.511,0.000){0}{\usebox{\plotpoint}}
\multiput(1129,868)(41.511,0.000){0}{\usebox{\plotpoint}}
\multiput(1141,868)(41.511,0.000){0}{\usebox{\plotpoint}}
\put(1156.46,867.73){\usebox{\plotpoint}}
\multiput(1166,867)(41.511,0.000){0}{\usebox{\plotpoint}}
\multiput(1178,867)(41.511,0.000){0}{\usebox{\plotpoint}}
\put(1197.92,866.39){\usebox{\plotpoint}}
\multiput(1203,866)(41.368,-3.447){0}{\usebox{\plotpoint}}
\multiput(1215,865)(41.368,-3.447){0}{\usebox{\plotpoint}}
\multiput(1227,864)(41.368,-3.447){0}{\usebox{\plotpoint}}
\put(1239.29,862.98){\usebox{\plotpoint}}
\multiput(1252,862)(40.946,-6.824){0}{\usebox{\plotpoint}}
\multiput(1264,860)(40.272,-10.068){0}{\usebox{\plotpoint}}
\put(1280.12,856.05){\usebox{\plotpoint}}
\multiput(1289,854)(38.318,-15.966){0}{\usebox{\plotpoint}}
\multiput(1301,849)(35.856,-20.916){0}{\usebox{\plotpoint}}
\put(1317.23,838.48){\usebox{\plotpoint}}
\multiput(1325,832)(27.187,-31.369){0}{\usebox{\plotpoint}}
\put(1343.00,806.59){\usebox{\plotpoint}}
\put(1356.53,767.53){\usebox{\plotpoint}}
\multiput(1362,747)(5.747,-41.111){2}{\usebox{\plotpoint}}
\multiput(1375,654)(2.163,-41.455){6}{\usebox{\plotpoint}}
\multiput(1387,424)(0.667,-41.506){7}{\usebox{\plotpoint}}
\put(1392,113){\usebox{\plotpoint}}
\end{picture}
        \begin{caption}
{Upper and lower bounds for the ionization probability for the 
$\psi_{100}$-state of the hydrogen atom in the first cycle of an
applied field $E_z=E_0\cos(1.5 t)$ for $ E_0=5,10,20$, when
neglecting the ``spreading of the wave''.}
        \end{caption}
    \end{center} 
\end{figure}

Figure 1 shows a plot of one cycle, when neglecting the term which is
independent of the electric field strength, i.e. the term $2 \tau$,
in the upper (3.32) and lower (3.33) bound. 
At the far ends one observes the curves
for the upper bounds, which approach the vertical for increasing $E_0$.
That is starting from the outside and going inwards the solid line $\equiv 
E_0 =20$, dotted line $\equiv E_0 =10$ and the next solid line $\equiv 
E_0 =5$. Next we have the lower bounds for $E_0=20 \equiv$ dotted line,
$E_0 =10 \equiv$ solid line and $E_0 =5 \equiv$ dotted line.  
We have used the same values as in \cite{Gelt1} and compare with figure
3 therein. Figure 1 clearly reproduces the features of Geltman's
results, indicating the dips at the half cycles and producing an
increasing ionization probability for increasing field strength. We
do, however, not observe any crossing for different field intensities.
When including  the $2 \tau$-term, this pattern will be moved above 
the trivial bound 1.
This may be avoided when achieving a better estimate for the factor
in front of $\tau$, for instance when integrating explicitly 
(\ref{3.7}) for a given pulse \cite{inprep}.
Figure 2 shows the  upper bound for four cycles and reproduces the
well known oscillatory behaviour superimposed by a spreading of the 
wave-packet of the Gordon-Volkov solution, the so-called 
over-the-barrier ionization.
\begin{figure}[h]
   \begin{center}
\setlength{\unitlength}{0.240900pt}
\ifx\plotpoint\undefined\newsavebox{\plotpoint}\fi
\sbox{\plotpoint}{\rule[-0.200pt]{0.400pt}{0.400pt}}%
\begin{picture}(1500,900)(0,0)
\font\gnuplot=cmr10 at 10pt
\gnuplot
\sbox{\plotpoint}{\rule[-0.200pt]{0.400pt}{0.400pt}}%
\put(220.0,113.0){\rule[-0.200pt]{292.934pt}{0.400pt}}
\put(220.0,113.0){\rule[-0.200pt]{0.400pt}{184.048pt}}
\put(220.0,113.0){\rule[-0.200pt]{4.818pt}{0.400pt}}
\put(198,113){\makebox(0,0)[r]{0}}
\put(1416.0,113.0){\rule[-0.200pt]{4.818pt}{0.400pt}}
\put(220.0,266.0){\rule[-0.200pt]{4.818pt}{0.400pt}}
\put(198,266){\makebox(0,0)[r]{0.2}}
\put(1416.0,266.0){\rule[-0.200pt]{4.818pt}{0.400pt}}
\put(220.0,419.0){\rule[-0.200pt]{4.818pt}{0.400pt}}
\put(198,419){\makebox(0,0)[r]{0.4}}
\put(1416.0,419.0){\rule[-0.200pt]{4.818pt}{0.400pt}}
\put(220.0,571.0){\rule[-0.200pt]{4.818pt}{0.400pt}}
\put(198,571){\makebox(0,0)[r]{0.6}}
\put(1416.0,571.0){\rule[-0.200pt]{4.818pt}{0.400pt}}
\put(220.0,724.0){\rule[-0.200pt]{4.818pt}{0.400pt}}
\put(198,724){\makebox(0,0)[r]{0.8}}
\put(1416.0,724.0){\rule[-0.200pt]{4.818pt}{0.400pt}}
\put(220.0,877.0){\rule[-0.200pt]{4.818pt}{0.400pt}}
\put(198,877){\makebox(0,0)[r]{1}}
\put(1416.0,877.0){\rule[-0.200pt]{4.818pt}{0.400pt}}
\put(220.0,113.0){\rule[-0.200pt]{0.400pt}{4.818pt}}
\put(220,68){\makebox(0,0){0}}
\put(220.0,857.0){\rule[-0.200pt]{0.400pt}{4.818pt}}
\put(342.0,113.0){\rule[-0.200pt]{0.400pt}{4.818pt}}
\put(342,68){\makebox(0,0){0.05}}
\put(342.0,857.0){\rule[-0.200pt]{0.400pt}{4.818pt}}
\put(463.0,113.0){\rule[-0.200pt]{0.400pt}{4.818pt}}
\put(463,68){\makebox(0,0){0.1}}
\put(463.0,857.0){\rule[-0.200pt]{0.400pt}{4.818pt}}
\put(585.0,113.0){\rule[-0.200pt]{0.400pt}{4.818pt}}
\put(585,68){\makebox(0,0){0.15}}
\put(585.0,857.0){\rule[-0.200pt]{0.400pt}{4.818pt}}
\put(706.0,113.0){\rule[-0.200pt]{0.400pt}{4.818pt}}
\put(706,68){\makebox(0,0){0.2}}
\put(706.0,857.0){\rule[-0.200pt]{0.400pt}{4.818pt}}
\put(828.0,113.0){\rule[-0.200pt]{0.400pt}{4.818pt}}
\put(828,68){\makebox(0,0){0.25}}
\put(828.0,857.0){\rule[-0.200pt]{0.400pt}{4.818pt}}
\put(950.0,113.0){\rule[-0.200pt]{0.400pt}{4.818pt}}
\put(950,68){\makebox(0,0){0.3}}
\put(950.0,857.0){\rule[-0.200pt]{0.400pt}{4.818pt}}
\put(1071.0,113.0){\rule[-0.200pt]{0.400pt}{4.818pt}}
\put(1071,68){\makebox(0,0){0.35}}
\put(1071.0,857.0){\rule[-0.200pt]{0.400pt}{4.818pt}}
\put(1193.0,113.0){\rule[-0.200pt]{0.400pt}{4.818pt}}
\put(1193,68){\makebox(0,0){0.4}}
\put(1193.0,857.0){\rule[-0.200pt]{0.400pt}{4.818pt}}
\put(1314.0,113.0){\rule[-0.200pt]{0.400pt}{4.818pt}}
\put(1314,68){\makebox(0,0){0.45}}
\put(1314.0,857.0){\rule[-0.200pt]{0.400pt}{4.818pt}}
\put(1436.0,113.0){\rule[-0.200pt]{0.400pt}{4.818pt}}
\put(1436,68){\makebox(0,0){0.5}}
\put(1436.0,857.0){\rule[-0.200pt]{0.400pt}{4.818pt}}
\put(220.0,113.0){\rule[-0.200pt]{292.934pt}{0.400pt}}
\put(1436.0,113.0){\rule[-0.200pt]{0.400pt}{184.048pt}}
\put(220.0,877.0){\rule[-0.200pt]{292.934pt}{0.400pt}}
\put(45,495){\makebox(0,0){I}}
\put(828,23){\makebox(0,0){time}}
\put(220.0,113.0){\rule[-0.200pt]{0.400pt}{184.048pt}}
\put(220,113){\usebox{\plotpoint}}
\multiput(220.00,113.61)(2.472,0.447){3}{\rule{1.700pt}{0.108pt}}
\multiput(220.00,112.17)(8.472,3.000){2}{\rule{0.850pt}{0.400pt}}
\multiput(232.00,116.59)(0.824,0.488){13}{\rule{0.750pt}{0.117pt}}
\multiput(232.00,115.17)(11.443,8.000){2}{\rule{0.375pt}{0.400pt}}
\multiput(245.00,124.58)(0.543,0.492){19}{\rule{0.536pt}{0.118pt}}
\multiput(245.00,123.17)(10.887,11.000){2}{\rule{0.268pt}{0.400pt}}
\multiput(257.58,135.00)(0.492,0.539){21}{\rule{0.119pt}{0.533pt}}
\multiput(256.17,135.00)(12.000,11.893){2}{\rule{0.400pt}{0.267pt}}
\multiput(269.00,148.58)(0.496,0.492){21}{\rule{0.500pt}{0.119pt}}
\multiput(269.00,147.17)(10.962,12.000){2}{\rule{0.250pt}{0.400pt}}
\multiput(281.00,160.59)(0.728,0.489){15}{\rule{0.678pt}{0.118pt}}
\multiput(281.00,159.17)(11.593,9.000){2}{\rule{0.339pt}{0.400pt}}
\multiput(294.00,169.61)(2.472,0.447){3}{\rule{1.700pt}{0.108pt}}
\multiput(294.00,168.17)(8.472,3.000){2}{\rule{0.850pt}{0.400pt}}
\put(306,170.67){\rule{2.891pt}{0.400pt}}
\multiput(306.00,171.17)(6.000,-1.000){2}{\rule{1.445pt}{0.400pt}}
\multiput(318.00,169.93)(1.123,-0.482){9}{\rule{0.967pt}{0.116pt}}
\multiput(318.00,170.17)(10.994,-6.000){2}{\rule{0.483pt}{0.400pt}}
\multiput(331.00,163.93)(0.669,-0.489){15}{\rule{0.633pt}{0.118pt}}
\multiput(331.00,164.17)(10.685,-9.000){2}{\rule{0.317pt}{0.400pt}}
\multiput(343.00,154.92)(0.600,-0.491){17}{\rule{0.580pt}{0.118pt}}
\multiput(343.00,155.17)(10.796,-10.000){2}{\rule{0.290pt}{0.400pt}}
\multiput(355.00,144.92)(0.600,-0.491){17}{\rule{0.580pt}{0.118pt}}
\multiput(355.00,145.17)(10.796,-10.000){2}{\rule{0.290pt}{0.400pt}}
\multiput(367.00,136.60)(1.797,0.468){5}{\rule{1.400pt}{0.113pt}}
\multiput(367.00,135.17)(10.094,4.000){2}{\rule{0.700pt}{0.400pt}}
\multiput(380.58,140.00)(0.492,0.712){21}{\rule{0.119pt}{0.667pt}}
\multiput(379.17,140.00)(12.000,15.616){2}{\rule{0.400pt}{0.333pt}}
\multiput(392.58,157.00)(0.492,0.841){21}{\rule{0.119pt}{0.767pt}}
\multiput(391.17,157.00)(12.000,18.409){2}{\rule{0.400pt}{0.383pt}}
\multiput(404.58,177.00)(0.493,0.774){23}{\rule{0.119pt}{0.715pt}}
\multiput(403.17,177.00)(13.000,18.515){2}{\rule{0.400pt}{0.358pt}}
\multiput(417.58,197.00)(0.492,0.755){21}{\rule{0.119pt}{0.700pt}}
\multiput(416.17,197.00)(12.000,16.547){2}{\rule{0.400pt}{0.350pt}}
\multiput(429.58,215.00)(0.492,0.539){21}{\rule{0.119pt}{0.533pt}}
\multiput(428.17,215.00)(12.000,11.893){2}{\rule{0.400pt}{0.267pt}}
\multiput(441.00,228.59)(1.033,0.482){9}{\rule{0.900pt}{0.116pt}}
\multiput(441.00,227.17)(10.132,6.000){2}{\rule{0.450pt}{0.400pt}}
\put(453,232.67){\rule{3.132pt}{0.400pt}}
\multiput(453.00,233.17)(6.500,-1.000){2}{\rule{1.566pt}{0.400pt}}
\multiput(466.00,231.93)(0.669,-0.489){15}{\rule{0.633pt}{0.118pt}}
\multiput(466.00,232.17)(10.685,-9.000){2}{\rule{0.317pt}{0.400pt}}
\multiput(478.58,221.79)(0.492,-0.539){21}{\rule{0.119pt}{0.533pt}}
\multiput(477.17,222.89)(12.000,-11.893){2}{\rule{0.400pt}{0.267pt}}
\multiput(490.58,208.41)(0.493,-0.655){23}{\rule{0.119pt}{0.623pt}}
\multiput(489.17,209.71)(13.000,-15.707){2}{\rule{0.400pt}{0.312pt}}
\multiput(503.58,191.09)(0.492,-0.755){21}{\rule{0.119pt}{0.700pt}}
\multiput(502.17,192.55)(12.000,-16.547){2}{\rule{0.400pt}{0.350pt}}
\multiput(515.00,174.92)(0.496,-0.492){21}{\rule{0.500pt}{0.119pt}}
\multiput(515.00,175.17)(10.962,-12.000){2}{\rule{0.250pt}{0.400pt}}
\multiput(527.58,164.00)(0.492,1.142){21}{\rule{0.119pt}{1.000pt}}
\multiput(526.17,164.00)(12.000,24.924){2}{\rule{0.400pt}{0.500pt}}
\multiput(539.58,191.00)(0.493,1.171){23}{\rule{0.119pt}{1.023pt}}
\multiput(538.17,191.00)(13.000,27.877){2}{\rule{0.400pt}{0.512pt}}
\multiput(552.58,221.00)(0.492,1.315){21}{\rule{0.119pt}{1.133pt}}
\multiput(551.17,221.00)(12.000,28.648){2}{\rule{0.400pt}{0.567pt}}
\multiput(564.58,252.00)(0.492,1.229){21}{\rule{0.119pt}{1.067pt}}
\multiput(563.17,252.00)(12.000,26.786){2}{\rule{0.400pt}{0.533pt}}
\multiput(576.58,281.00)(0.492,0.970){21}{\rule{0.119pt}{0.867pt}}
\multiput(575.17,281.00)(12.000,21.201){2}{\rule{0.400pt}{0.433pt}}
\multiput(588.58,304.00)(0.493,0.576){23}{\rule{0.119pt}{0.562pt}}
\multiput(587.17,304.00)(13.000,13.834){2}{\rule{0.400pt}{0.281pt}}
\multiput(601.00,319.59)(1.267,0.477){7}{\rule{1.060pt}{0.115pt}}
\multiput(601.00,318.17)(9.800,5.000){2}{\rule{0.530pt}{0.400pt}}
\multiput(613.00,322.93)(1.267,-0.477){7}{\rule{1.060pt}{0.115pt}}
\multiput(613.00,323.17)(9.800,-5.000){2}{\rule{0.530pt}{0.400pt}}
\multiput(625.00,317.92)(0.497,-0.493){23}{\rule{0.500pt}{0.119pt}}
\multiput(625.00,318.17)(11.962,-13.000){2}{\rule{0.250pt}{0.400pt}}
\multiput(638.58,302.82)(0.492,-0.841){21}{\rule{0.119pt}{0.767pt}}
\multiput(637.17,304.41)(12.000,-18.409){2}{\rule{0.400pt}{0.383pt}}
\multiput(650.58,282.40)(0.492,-0.970){21}{\rule{0.119pt}{0.867pt}}
\multiput(649.17,284.20)(12.000,-21.201){2}{\rule{0.400pt}{0.433pt}}
\multiput(662.58,259.13)(0.492,-1.056){21}{\rule{0.119pt}{0.933pt}}
\multiput(661.17,261.06)(12.000,-23.063){2}{\rule{0.400pt}{0.467pt}}
\multiput(674.58,238.00)(0.493,0.655){23}{\rule{0.119pt}{0.623pt}}
\multiput(673.17,238.00)(13.000,15.707){2}{\rule{0.400pt}{0.312pt}}
\multiput(687.58,255.00)(0.492,1.703){21}{\rule{0.119pt}{1.433pt}}
\multiput(686.17,255.00)(12.000,37.025){2}{\rule{0.400pt}{0.717pt}}
\multiput(699.58,295.00)(0.492,1.746){21}{\rule{0.119pt}{1.467pt}}
\multiput(698.17,295.00)(12.000,37.956){2}{\rule{0.400pt}{0.733pt}}
\multiput(711.58,336.00)(0.493,1.448){23}{\rule{0.119pt}{1.238pt}}
\multiput(710.17,336.00)(13.000,34.430){2}{\rule{0.400pt}{0.619pt}}
\multiput(724.58,373.00)(0.492,1.315){21}{\rule{0.119pt}{1.133pt}}
\multiput(723.17,373.00)(12.000,28.648){2}{\rule{0.400pt}{0.567pt}}
\multiput(736.58,404.00)(0.492,0.884){21}{\rule{0.119pt}{0.800pt}}
\multiput(735.17,404.00)(12.000,19.340){2}{\rule{0.400pt}{0.400pt}}
\multiput(748.00,425.59)(0.669,0.489){15}{\rule{0.633pt}{0.118pt}}
\multiput(748.00,424.17)(10.685,9.000){2}{\rule{0.317pt}{0.400pt}}
\multiput(760.00,432.94)(1.797,-0.468){5}{\rule{1.400pt}{0.113pt}}
\multiput(760.00,433.17)(10.094,-4.000){2}{\rule{0.700pt}{0.400pt}}
\multiput(773.58,427.65)(0.492,-0.582){21}{\rule{0.119pt}{0.567pt}}
\multiput(772.17,428.82)(12.000,-12.824){2}{\rule{0.400pt}{0.283pt}}
\multiput(785.58,412.26)(0.492,-1.013){21}{\rule{0.119pt}{0.900pt}}
\multiput(784.17,414.13)(12.000,-22.132){2}{\rule{0.400pt}{0.450pt}}
\multiput(797.58,387.88)(0.493,-1.131){23}{\rule{0.119pt}{0.992pt}}
\multiput(796.17,389.94)(13.000,-26.940){2}{\rule{0.400pt}{0.496pt}}
\multiput(810.58,358.02)(0.492,-1.401){21}{\rule{0.119pt}{1.200pt}}
\multiput(809.17,360.51)(12.000,-30.509){2}{\rule{0.400pt}{0.600pt}}
\multiput(822.58,327.79)(0.492,-0.539){21}{\rule{0.119pt}{0.533pt}}
\multiput(821.17,328.89)(12.000,-11.893){2}{\rule{0.400pt}{0.267pt}}
\multiput(834.58,317.00)(0.492,2.133){21}{\rule{0.119pt}{1.767pt}}
\multiput(833.17,317.00)(12.000,46.333){2}{\rule{0.400pt}{0.883pt}}
\multiput(846.58,367.00)(0.493,2.043){23}{\rule{0.119pt}{1.700pt}}
\multiput(845.17,367.00)(13.000,48.472){2}{\rule{0.400pt}{0.850pt}}
\multiput(859.58,419.00)(0.492,2.133){21}{\rule{0.119pt}{1.767pt}}
\multiput(858.17,419.00)(12.000,46.333){2}{\rule{0.400pt}{0.883pt}}
\multiput(871.58,469.00)(0.492,1.875){21}{\rule{0.119pt}{1.567pt}}
\multiput(870.17,469.00)(12.000,40.748){2}{\rule{0.400pt}{0.783pt}}
\multiput(883.58,513.00)(0.493,1.290){23}{\rule{0.119pt}{1.115pt}}
\multiput(882.17,513.00)(13.000,30.685){2}{\rule{0.400pt}{0.558pt}}
\multiput(896.58,546.00)(0.492,0.884){21}{\rule{0.119pt}{0.800pt}}
\multiput(895.17,546.00)(12.000,19.340){2}{\rule{0.400pt}{0.400pt}}
\multiput(908.00,567.59)(1.267,0.477){7}{\rule{1.060pt}{0.115pt}}
\multiput(908.00,566.17)(9.800,5.000){2}{\rule{0.530pt}{0.400pt}}
\multiput(920.00,570.93)(0.669,-0.489){15}{\rule{0.633pt}{0.118pt}}
\multiput(920.00,571.17)(10.685,-9.000){2}{\rule{0.317pt}{0.400pt}}
\multiput(932.58,559.90)(0.493,-0.814){23}{\rule{0.119pt}{0.746pt}}
\multiput(931.17,561.45)(13.000,-19.451){2}{\rule{0.400pt}{0.373pt}}
\multiput(945.58,537.16)(0.492,-1.358){21}{\rule{0.119pt}{1.167pt}}
\multiput(944.17,539.58)(12.000,-29.579){2}{\rule{0.400pt}{0.583pt}}
\multiput(957.58,504.47)(0.492,-1.573){21}{\rule{0.119pt}{1.333pt}}
\multiput(956.17,507.23)(12.000,-34.233){2}{\rule{0.400pt}{0.667pt}}
\multiput(969.58,467.48)(0.493,-1.567){23}{\rule{0.119pt}{1.331pt}}
\multiput(968.17,470.24)(13.000,-37.238){2}{\rule{0.400pt}{0.665pt}}
\multiput(982.58,433.00)(0.492,1.703){21}{\rule{0.119pt}{1.433pt}}
\multiput(981.17,433.00)(12.000,37.025){2}{\rule{0.400pt}{0.717pt}}
\multiput(994.58,473.00)(0.492,2.650){21}{\rule{0.119pt}{2.167pt}}
\multiput(993.17,473.00)(12.000,57.503){2}{\rule{0.400pt}{1.083pt}}
\multiput(1006.58,535.00)(0.492,2.607){21}{\rule{0.119pt}{2.133pt}}
\multiput(1005.17,535.00)(12.000,56.572){2}{\rule{0.400pt}{1.067pt}}
\multiput(1018.58,596.00)(0.493,2.083){23}{\rule{0.119pt}{1.731pt}}
\multiput(1017.17,596.00)(13.000,49.408){2}{\rule{0.400pt}{0.865pt}}
\multiput(1031.58,649.00)(0.492,1.832){21}{\rule{0.119pt}{1.533pt}}
\multiput(1030.17,649.00)(12.000,39.817){2}{\rule{0.400pt}{0.767pt}}
\multiput(1043.58,692.00)(0.492,1.142){21}{\rule{0.119pt}{1.000pt}}
\multiput(1042.17,692.00)(12.000,24.924){2}{\rule{0.400pt}{0.500pt}}
\multiput(1055.00,719.58)(0.590,0.492){19}{\rule{0.573pt}{0.118pt}}
\multiput(1055.00,718.17)(11.811,11.000){2}{\rule{0.286pt}{0.400pt}}
\multiput(1068.00,728.93)(1.033,-0.482){9}{\rule{0.900pt}{0.116pt}}
\multiput(1068.00,729.17)(10.132,-6.000){2}{\rule{0.450pt}{0.400pt}}
\multiput(1080.58,720.54)(0.492,-0.927){21}{\rule{0.119pt}{0.833pt}}
\multiput(1079.17,722.27)(12.000,-20.270){2}{\rule{0.400pt}{0.417pt}}
\multiput(1092.58,696.88)(0.492,-1.444){21}{\rule{0.119pt}{1.233pt}}
\multiput(1091.17,699.44)(12.000,-31.440){2}{\rule{0.400pt}{0.617pt}}
\multiput(1104.58,662.09)(0.493,-1.686){23}{\rule{0.119pt}{1.423pt}}
\multiput(1103.17,665.05)(13.000,-40.046){2}{\rule{0.400pt}{0.712pt}}
\multiput(1117.58,618.08)(0.492,-2.004){21}{\rule{0.119pt}{1.667pt}}
\multiput(1116.17,621.54)(12.000,-43.541){2}{\rule{0.400pt}{0.833pt}}
\multiput(1129.00,576.93)(1.267,-0.477){7}{\rule{1.060pt}{0.115pt}}
\multiput(1129.00,577.17)(9.800,-5.000){2}{\rule{0.530pt}{0.400pt}}
\multiput(1141.58,573.00)(0.492,3.124){21}{\rule{0.119pt}{2.533pt}}
\multiput(1140.17,573.00)(12.000,67.742){2}{\rule{0.400pt}{1.267pt}}
\multiput(1153.58,646.00)(0.493,2.875){23}{\rule{0.119pt}{2.346pt}}
\multiput(1152.17,646.00)(13.000,68.130){2}{\rule{0.400pt}{1.173pt}}
\multiput(1166.58,719.00)(0.492,2.909){21}{\rule{0.119pt}{2.367pt}}
\multiput(1165.17,719.00)(12.000,63.088){2}{\rule{0.400pt}{1.183pt}}
\multiput(1178.58,787.00)(0.492,2.478){21}{\rule{0.119pt}{2.033pt}}
\multiput(1177.17,787.00)(12.000,53.780){2}{\rule{0.400pt}{1.017pt}}
\multiput(1190.58,845.00)(0.491,1.642){17}{\rule{0.118pt}{1.380pt}}
\multiput(1189.17,845.00)(10.000,29.136){2}{\rule{0.400pt}{0.690pt}}
\put(1250.17,872){\rule{0.400pt}{1.100pt}}
\multiput(1249.17,874.72)(2.000,-2.717){2}{\rule{0.400pt}{0.550pt}}
\multiput(1252.58,865.50)(0.492,-1.875){21}{\rule{0.119pt}{1.567pt}}
\multiput(1251.17,868.75)(12.000,-40.748){2}{\rule{0.400pt}{0.783pt}}
\multiput(1264.58,820.39)(0.492,-2.219){21}{\rule{0.119pt}{1.833pt}}
\multiput(1263.17,824.19)(12.000,-48.195){2}{\rule{0.400pt}{0.917pt}}
\multiput(1276.58,768.56)(0.493,-2.162){23}{\rule{0.119pt}{1.792pt}}
\multiput(1275.17,772.28)(13.000,-51.280){2}{\rule{0.400pt}{0.896pt}}
\multiput(1289.58,721.00)(0.492,3.081){21}{\rule{0.119pt}{2.500pt}}
\multiput(1288.17,721.00)(12.000,66.811){2}{\rule{0.400pt}{1.250pt}}
\multiput(1301.58,793.00)(0.492,3.598){21}{\rule{0.119pt}{2.900pt}}
\multiput(1300.17,793.00)(12.000,77.981){2}{\rule{0.400pt}{1.450pt}}
\end{picture}
        \begin{caption}
{Upper bound for the ionization probability for the
$\psi_{100}$-state of the hydrogen atom in the first four cycles
of an applied field $E_z=10 \cos(50 t)$.}
        \end{caption}
    \end{center} 
\end{figure}

\section{Conclusions}
In conclusion we can say that, according to our arguments atoms do not
become resistant to ionization when exposed to short ultra-intense
laser pulses. We therefore disagree with the opposite point of view,
which is sustained through
numerous numerical studies partly based on explicit solutions of the
Schr\"odinger equation and partly based on perturbative methods.
In particular we have commented above on the problems of the latter
methods. It is not the intention of this paper to discuss the problems
of numerical methods, but we
like to remark that those studies are in general very complex and
subject to many possible errors which are difficult to check for
second parties. We think that the virtue of our arguments is that
they are analytic and transparent to the reader. 
An extension of our results to multi-particle systems (thus including
atoms and molecules with several electrons and not necessarily electrically
neutral) may be found in \cite{KS2}.
Needless to say, since our results
are  of a qualitative nature, in the sense that they merely provide
bounds and that, since there are no
explicit solutions for the Schr\"odinger equation available, for
precise predictions of ionization rates one needs more numerical data.

\par
\noindent
{\bf Acknowledgment}: 
We would like to thank J. Burgd\"orfer, J.H. Eberly, S. Geltman and  
W. Sandner for discussions and correspondences. 
A.F. is grateful to C. Figueira de Morisson Faria
for useful discussions and to the International School for Advanced
Studies (Trieste, Italy) for its kind hospitality.

\appendix 
\section*{Appendix A}
\renewcommand{\theequation}{\mbox{{A}.\arabic{equation}}}
\setcounter{equation}{0}

The aim of this appendix is to prove the following bound for the Coulomb potential
$V(\vec{x}) =-{1\over r} ( r=|\vec{x}| ) $ on $L^2 ( \Mrr^3, d^3x)=L^2$.\\
\\{\bf Lemma A:} {\it The following operator norm bound holds }
\begin{equation}\label{A1}
\left\|{1\over r} (-\Delta +{\bf 1})^{-1} \right\|\le 6.35
\end{equation}
The proof optimizes well known a priori bounds which we take from  \cite{RS}:

Let $L^\infty$ be the space of all Lebesgue measurable functions $\varphi$ on
$\Mrr^3$  such that $|\varphi (\vec{x}) |\le M<\infty$ almost everywhere. 
The smallest such $M$ is
denoted by $\|\varphi\|_\infty$. Then one has the a priori estimate 
(\cite{RS} p.56)
\begin{equation}\label{A2}
\|\varphi\|_\infty \le a(\rho) \| -\Delta\varphi\|_2 +b(\rho) \|\varphi\|_2
\end{equation}
where $\|~~~\|_2$ denotes the $L^2$ -norm, i.e. 
$\| \varphi \|_2 = \int | \varphi(\vec{x})^2 | d^3 x $  . Here
\begin{eqnarray}\label{A3}
a(\rho) &=& c\rho^{-1}\nonumber\\
b(\rho) &=& c\rho^3
\end{eqnarray}
with
\begin{equation}\label{A4}
c = \left( \int\limits^{+\infty}_{-\infty} {1\over (1 +\lambda^2)^2} 
d\lambda\right)
^{1\over 2} =\sqrt{\pi\over 2}
\end{equation}
and $\rho>0$ may be chosen arbitrary.

Let $R>0$ be arbitrary and write
$$
{1\over r} =V_1^R + V_2^R
$$
with
\begin{eqnarray*}
V_1^R(r) &=& {\theta (r<R)\over r}\\
V_2^R(r) &=& {\theta(r\ge R)\over r} .
\end{eqnarray*}
Then one has the a priori bound (see \cite{RS} p.165)
\begin{equation}\label{A5}
\left\|{1\over r} \varphi \right\|_2 \le a(\rho) \;\|V_1^R\|_2~~\| 
-\Delta \varphi\|_2 +
\left( b(\rho) +\| V_2^R\|_\infty \right) \|\varphi\|_2.
\end{equation}
Since the estimates $\|\varphi\|_2 \le \| (-\Delta +{\bf 1})\varphi\|_2, \| -\Delta
\varphi \|_2 \le\| (-\Delta +{\bf 1})\varphi\|_2$ are trivially valid,
this gives
\begin{equation}\label{A6}
\left\| {1\over r} (-\Delta +{\bf 1})^{-1} \right\| \le a(\rho) \| 
V_1^R\|_2 +b(\rho)+\|V_2^R\|_\infty .
\end{equation}
Obviously
\begin{eqnarray}\label{A7}
&&\|V_1^R\|_2 = (4\pi R)^{1\over 2}, \nonumber\\
&&\| V_2^R\|_\infty = {1\over R}.
\end{eqnarray}
Inserting (A.3), (A.4) and (A.7) into (A.6) gives 
\begin{equation}\label{A8}
\left\|{1\over r} (-\Delta+{\bf 1})^{-1} \right\| 
\le \pi\sqrt2\rho^{-1} R^{1\over 2} +\sqrt{\pi\over 2}
\rho^3 +{1\over R}.
\end{equation}
for all $\rho>0,R>0$. The claim now follows by optimizing w.r.t. $\rho$ and $R$.
A short calculation finally gives
\begin{equation}\label{A9}
\left\|{1\over r} (-\Delta +{\bf 1} )^{-1} \right\| 
\le 11 \cdot {\pi^{7\over 11} \over
2^ {6\over 11} 3^{9\over 11} }
\end{equation}
which is (A1).

\appendix 
\section*{Appendix B}
\renewcommand{\theequation}{\mbox{{B}.\arabic{equation}}}
\setcounter{equation}{0}

In this section we will study the quantity  $\langle \psi| V(\vec{x}-\vec{y})^k |\psi\rangle$
for $k=1$ and 2 where $V(\vec{x}) =-{1\over|\vec{x}|}(\vec{x}\in 
\Mrr^3)$ is the Coulomb
potential and where $\psi$ is the normalized ground state wave  function 
$\psi_{100}$ for the hydrogen atom, which is rotationally invariant. 
Therefore this quantity depends on $|\vec{y}|$ only.

\noindent{\bf Lemma B:} {\it Both $-\langle \psi| V(\vec{x}-\vec{y})
|\psi\rangle$ 
 and $\langle \psi|V(\vec{x}-\vec{y})^2|\psi\rangle$ are  decreasing functions of
$|\vec{y}|$ for $\psi=\psi_{100}$.}

Intuitively this result is clear: $\psi(\vec{x})$ has its maximum at $\vec{x}=0$ and
$-V(\vec{x}-\vec{y})\ge 0$ has its singularity at $\vec{x}=\vec{y}$ so 
their overlap is maximal when $\vec{y}=0$.

Before we give a proof, we first establish an important consequence. Indeed we
claim that for $\psi=\psi_{100}$
\begin{equation}\label{B1}
\| (V(\vec{x}-\vec{y})-V(\vec{x})) \psi\| \le 2 
\end{equation}
for all $\vec{y}\in  \Mrr^3$. To see this we write 
\begin{eqnarray}\label{B2}
\| (V(\vec{x} -\vec{y}) -V(\vec{x}))\psi\|^2 
&=&\langle \psi,V(\vec{x} -\vec{y})^2 \psi\rangle
-2 \langle\psi, V(\vec{x} -\vec{y}) V(\vec{x})\psi \rangle
+\langle \psi,V(\vec{x})^2 \psi \rangle\nonumber\\
&\le&\langle \psi,V(\vec{x} -\vec{y})^2 \psi \rangle +\langle\psi,V(\vec{x})^2\psi\rangle
\end{eqnarray}
since $V(\vec{x} -\vec{y}) V(\vec{x}) \ge0$ as an operator. By Lemma B, the right hand side
of (B.2) takes its maximum at $\vec{y}=0$ proving the claim since
$\langle\psi| {1\over|\vec{x}|^2}|\psi \rangle =2$ (see e.g. \cite{LL}).

To prove the lemma, it suffices to consider $\vec{y}$ to be of the form $\vec{y}=ce_z$ with
$c>0$. Now we use the well known formula
\begin{equation}\label{B3}
{1\over |\vec{x}-ce_z|} = {1\over r_>} \sum^\infty_{\ell =0} \left( {r_<\over r_>} \right)^\ell
P_\ell (\cos \vartheta)
\end{equation}
where $r_> = Max (r,c), r_< = Min (r,c)$ and where $(r=|\vec{x}|, \vartheta, \varphi)$
are the polar coordinates of $\vec{x}$. By the orthogonality relations of the Legendre
polynomials and since $\psi$ is the ground state we obtain
\begin{eqnarray}\label{B4}
-\langle\psi|V(\vec{x} -ce_z)|\psi\rangle &=& \langle \psi | {1\over r_>} | \psi \rangle\nonumber\\
&=& 4 \int^\infty_0 r^2 e^{-2r} {1\over r_>} dr \\
&=& {4\over c} \int^c_0 r^2 e^{-2r} dr + 4 \int^\infty_c r e^{-2r} dr.\nonumber
\end{eqnarray}
This function is differentiable in $c$ for $c>0$ and its derivative is easily
seen to be $\le 0$, proving the first claim. Next we have (again by the orthogonality
of the Legendre polynomials)
\begin{eqnarray}\label{B5}
\langle \psi|V(\vec{x} -ce_z)^2|\psi \rangle &=&\langle\psi| {1\over (r_>)^2}
\sum^\infty_{\ell=0} \left( {r_<\over r_>} \right)^{2\ell} {1\over 2\ell +1} 
|\psi \rangle \nonumber \\
&=& {4\over c^2} \int_0^c r^2 e^{-2r} \sum^\infty_{\ell=0} \left({r\over c}
\right)^{2\ell}
{1\over 2\ell +1} dr \\
\;\;\;\;\;\;\; &&+ 4 \int^\infty_c e^{-2r} \sum^\infty_{\ell=0} 
\left( {c\over r}\right) ^{2\ell} {1\over 2\ell +1} dr.\nonumber
\end{eqnarray}
Now for $0<x<1$ we have 
\begin{equation}\label{B6}
\sum^\infty _{\ell=0} x^{2\ell} {1\over 2\ell +1} = {1\over x} \sum^\infty_{\ell=0}
{x^{2\ell+1}\over 2\ell+1} ={1\over 2x} (\ln(1+x)-\ln(1-x)).
\end{equation}
Inserting (B6) into (B5) gives
\begin{eqnarray}\label {B7}
\langle\psi| V(\vec{x} -ce_z)^2|\psi \rangle &=& {2\over c}\int^c_0 r e^{-2r}
\left(\ln \left(1+{r\over c}\right)-\ln\left(1-{r\over c}\right)
\right)dr\nonumber\\
&& + {2\over c} \int^\infty_c r e^{-2r} \left(\ln\left(1+{c\over r}\right)
 -\ln\left(1-{c\over r}\right) \right) dr.
\end{eqnarray}
Now the right hand side is not differentiable in $c$. To remedy this we regularize
and consider the quantity $(0<\varepsilon <1)$
\begin{eqnarray}\label {B8}
0\le F(c,\varepsilon) ={G(c,\varepsilon)\over c}= {2\over c} 
\int_0^c r e^{-2r} \left[\ln(1+{r\over c})
-\ln(1-{r\over c} (1-\varepsilon))\right] dr\nonumber\\
+{2\over c} \int_c^\infty r e^{-2r} [\ln(1+{c\over r})-\ln
(1-{c\over r} (1-\varepsilon))] dr.
\end{eqnarray}
Since $\lim_{\varepsilon\to 0} F(c,\varepsilon)=\langle \psi|V(\vec{x} -ce_z)^2|\psi\rangle$
it suffices to show that
\begin{equation}\label{B9}
{d\over dc} F(c,\varepsilon) ={1\over c^2} (c{dG\over dc} (c,\varepsilon)
 -G(c,\varepsilon)) \le0
\end{equation}
for all $0<\varepsilon<1$, since then (B7) is also monotonically decreasing
in $c>0$. Now
\begin{eqnarray}\label{B10}
{d\over dc} G(c,\varepsilon) = 
&&-2\int^c_0 r e^{-2r} \left[ {r\over c^2}~ {1\over 1+{r\over c}}
+{r(1-\varepsilon)\over c^2}~ {1\over 1-{r\over c} (1-\varepsilon)} \right] dr\nonumber\\
&&+2 \int^\infty_c re^{-2r} \left[ {1\over r}~ {1\over 1+{c\over r}}
+{(1-\varepsilon)\over r}~ {1\over 1- {c\over r} (1-\varepsilon)} \right] dr.
\end{eqnarray}
The first integral on the r.h.s. of (B10) is negative. By (B8), (B9) and (B10) 
it therefore suffices
to show that for any $r>0$
\begin{equation}\label{B11}
{c\over r} {1\over 1 + {c\over r}} - \ln (1+{c\over r}) \le 0.
\end{equation}
and 
\begin{equation}\label{B12}
{c(1-\varepsilon) \over r} {1\over 1-{c\over r}(1-\varepsilon)} +\ln
\left(1 - {c\over r} (1-\varepsilon)\right) \le 0.
\end{equation}
Now
$$
\ln(1+{c\over r} ) ={1\over r} \int_0^c {1\over 1+{c'\over r}} dc' \ge
{1\over r} {1\over 1+{c\over r}} \cdot c
$$
which is (B11). (B12) is proved in the same fashion. This concludes the proof
of Lemma B.


\end{document}